%% file: main.tex
\documentclass[journal]{IEEEtran}
\PassOptionsToPackage{hyphens}{url}
\usepackage{hyperref}
\usepackage{graphicx}
\usepackage{footnote}
\usepackage{url}
\usepackage[table,xcdraw]{xcolor}
\usepackage[caption=false,font=footnotesize]{subfig}
\usepackage{multirow}
\usepackage{adjustbox}
\usepackage{pifont}
\usepackage{amsmath}
\usepackage{marvosym}
\usepackage{booktabs}
\usepackage{acronym}
\usepackage{tabu}
\usepackage{xfrac}
\usepackage[nice]{nicefrac}

\input{acronym}
\begin{document}
	\title{Hardware Implementation of Deep Network Accelerators Towards Healthcare and Biomedical Applications}
	\author{Mostafa Rahimi Azghadi\textsuperscript{\href{https://orcid.org/0000-0001-7975-3985}{\includegraphics[scale=0.035]{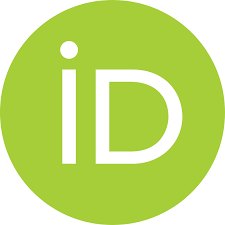}}},~\IEEEmembership{Senior Member,~IEEE,}
		Corey Lammie\textsuperscript{\href{https://orcid.org/0000-0001-5564-1356}{\includegraphics[scale=0.035]{ORCID.png}}},~\IEEEmembership{Student Member,~IEEE,}
		Jason~K.~Eshraghian\textsuperscript{\href{https://orcid.org/0000-0002-5832-4054}{\includegraphics[scale=0.035]{ORCID.png}}},~\IEEEmembership{Member,~IEEE,}
		Melika Payvand\textsuperscript{\href{https://orcid.org/0000-0001-5400-067X}{\includegraphics[scale=0.035]{ORCID.png}}},~\IEEEmembership{Member,~IEEE,}
		Elisa Donati\textsuperscript{\href{https://orcid.org/0000-0002-8091-1298}{\includegraphics[scale=0.035]{ORCID.png}}},~\IEEEmembership{Member,~IEEE,}		
		Bernab{\'e}~Linares-Barranco\textsuperscript{\href{https://orcid.org/0000-0002-1813-4889}{\includegraphics[scale=0.035]{ORCID.png}}},~\IEEEmembership{Fellow,~IEEE}
		and~Giacomo~Indiveri\textsuperscript{\href{https://orcid.org/0000-0002-7109-1689}{\includegraphics[scale=0.035]{ORCID.png}}},~\IEEEmembership{Senior Member,~IEEE}
		\thanks{M. Rahimi Azghadi and Corey Lammie are with the College of Science and Engineering, James Cook University, QLD 4811, Australia. e-mail: mostafa.rahimiazghadi@jcu.edu.au}
		\thanks{J. K. Eshraghian is with the Department of Electrical Engineering and Computer Science, The University of Michigan, Ann Arbor, MI 48109-2122, USA.}
		\thanks{M. Payvand, E. Donati, and G. Indiveri are with the Institute of Neuroinformatics, University and ETH Zurich, Switzerland.}
		\thanks{B. Linares-Barranco is with the Instituto de Microelectr{\'o}nica de Sevilla IMSE-CNM (CSIC and Universidad de Sevilla), Sevilla, Spain.\hspace{-1em}\rule{3cm}{0.5pt} \newline \textcopyright  \hspace{1pt} This work is licensed under a Creative Commons Attribution 4.0 License. For more information, see \url{https://creativecommons.org/licenses/by/4.0/}}}
	\markboth{ACCEPTED BY IEEE TRANSACTIONS ON BIOMEDICAL CIRCUITS AND SYSTEMS, 2020}%
	{Rahimi Azghadi \MakeLowercase{\textit{et al.}}: Neuromorphic Hardware for Biomedical Applications}
	\maketitle
	\begin{abstract}
          The advent of dedicated \ac{DL} accelerators and neuromorphic processors has brought on new opportunities for applying both Deep and \ac{SNN} algorithms to healthcare and biomedical applications at the edge. This can facilitate the advancement of medical \ac{IoT} systems and \ac{PoC} devices. In this paper, we provide a tutorial describing how various technologies including emerging memristive devices,  \acp{FPGA}, and Complementary Metal Oxide Semiconductor (CMOS) can be used to develop efficient \ac{DL} accelerators to solve a wide variety of diagnostic, pattern recognition, and signal processing problems in healthcare. Furthermore, we explore how spiking neuromorphic processors can complement their \ac{DL} counterparts for processing biomedical signals.   The tutorial is augmented with case studies of the vast literature on neural network and neuromorphic hardware as applied to the healthcare domain. We benchmark various hardware platforms by performing a sensor fusion signal processing task combining \ac{EMG} signals with computer vision. Comparisons are made between dedicated neuromorphic processors and embedded AI accelerators in terms of inference latency and energy. Finally, we provide our analysis of the field and share a perspective on the advantages, disadvantages, challenges, and opportunities that various accelerators and neuromorphic processors introduce to healthcare and biomedical domains. 
	\end{abstract}
	\IEEEpeerreviewmaketitle
	\section{Introduction}
\begin{figure*}[!t]
		\centering
		\includegraphics[width=1\textwidth]{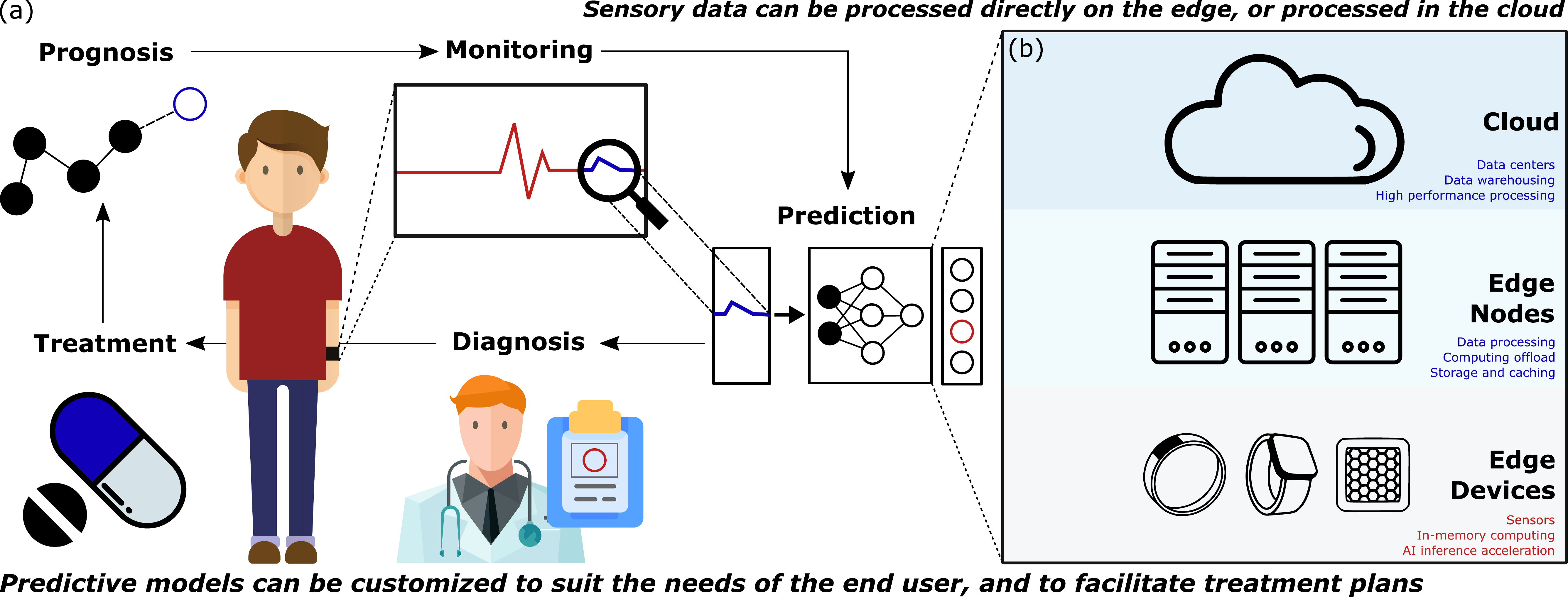}
		\caption{A depiction of (a) the usage of DL in a smart healthcare setting, which typically involves monitoring, prediction, diagnosis, treatment, and prognosis. The various parts of the DL-based healthcare system can run on (b) the three levels of the IoT, i.e. edge devices, edge nodes, and the cloud. However, for healthcare \ac{IoT} and \ac{PoC} processing, edge learning and inference is preferred.}
		\label{fig:DLhealth}
	\end{figure*}

    \acresetall
	\IEEEPARstart{A}{rtificial} intelligence is uniquely poised to cope with the growing demands of the universal healthcare system~\cite{Rong2020}. The healthcare industry is projected to reach over 10 trillion dollars by 2022, and the associated workload on medical practitioners is expected to grow concurrently~\cite{Arevalo2020}. As the reliability of \ac{DL} improves, it has pervaded various facets of healthcare from monitoring~\cite{Jindal2016,Sundaravadivel2018}, to prediction~\cite{Shi2018}, diagnosis~\cite{Liu2019}, treatment~\cite{Liu2019a}, and prognosis~\cite{Zhu2020}. Fig.~\ref{fig:DLhealth}(a) shows how data collected from the patient, which may be a combination of bio-samples, medical images, temperature, movement, etc., can be processed using a smart \ac{DL} system that monitors the patient for anomalies and/or to predict diseases. \ac{DL} systems can be used to recommend treatment options and prognosis, which further affect monitoring and prediction in a closed-loop scenario. 
	
The capacity of \ac{AI} to meet or exceed the performance of human experts in medical-data analysis \cite{mckinney2020international, hannun2019cardiologist, Esteva2017} can, in part, be attributed to the continued improvement of high-performance computing platforms such as \acp{GPU}~\cite{Kalaiselvi2017} and customized \ac{ML} hardware~\cite{Jouppi2017}. These can now process and learn from a large amount of multi-modal heterogeneous general and medical data~\cite{Esteva2019}. This was not readily achievable a decade ago. 
	
	While the field of \ac{DL} has been growing at an astonishing rate in terms of performance, network size, and training run time, the development of dedicated hardware to process \ac{DL} algorithms is struggling to keep up. Concretely, the compute loads of \ac{DL} have doubled every 3.4 months since 2012. Moore's Law targets the doubling of compute power every 18-24 months, and appears to be slowing down~\cite{perrault2019ai}. The progress in hardware accelerator development currently relies on advances by a handful of technology companies, most notably Nvidia and its \acp{GPU}~\cite{Peter2009,jia2018dissecting} and Google and its \acp{TPU}~\cite{Jouppi2017}, in addition to new startups and research groups developing \acp{ASIC} for \ac{DL} training and acceleration. While there are significant advances in tailoring deep network models and algorithms for various healthcare and biomedical applications~\cite{Zemouri2019}, most computationally expensive deep networks are trained on either \acp{GPU} or in data centers~\cite{Kalaiselvi2017,Smistad2015}. The latter typically requires access to cloud computing services which is not only costly and comes with high power demands, but also compromises data privacy. This is distinct to the effective deployment of \ac{DL} at the edge on an increasing number of medical \ac{IoT} devices~\cite{Farahani2020} and \ac{PoC} systems~\cite{Xie2019}, as illustrated in Fig.~\ref{fig:DLhealth}(b). Edge learning and inference enables the option to move processing away from the cloud. This is critical for highly sensitive medical data and offline operation. Edge-based processing must combine compactness, low-power, and rapid (high throughput) at a low-cost, to make smart health monitoring viable and affordable for integration into human life~\cite{Hartmann2019}.    
	
Specialized embedded \ac{DL} accelerators, such as the Nvidia Jetson and Xavier series~\cite{Azimi2017}, and the Movidius Neural Compute Stick~\cite{Sethi2018,Sahu2018}, have shown the promise of edge computing. More recently, the Nvidia Clara Embedded was released as a healthcare-specific edge accelerator. This is a computing platform for edge-enabled AI on the \ac{IoMT}. However, embedded devices remain relatively power hungry and costly, and many state-of-the-art algorithms far exceed the memory bandwidth of resource-constrained devices. They are not yet ideal learning/inference engines for ambient-assisted precision medicine systems. There is a need for innovative systems which can satisfy the stringent requirements of healthcare edge devices to be made affordable to the community at large scales. 
	
To that end, in this paper we focus on the use of three various hardware technologies to develop dedicated deep network accelerators which will be discussed from a biomedical and healthcare application point-of-view. The three technologies that we cover here are CMOS, memristors, and \acp{FPGA}. It is worth noting that, while our focus targets edge inference engines in the biomedical domain, the techniques and hardware advantages discussed here are likely to be useful for efficient offline deep network learning, or online on-chip learning. Herein, the term \ac{DL} `accelerator' is used to refer to a device that is able to perform \ac{DL} inference and potentially training.
    
    \begin{figure*}[!t]
		\centering
		\includegraphics[width=0.9\textwidth]{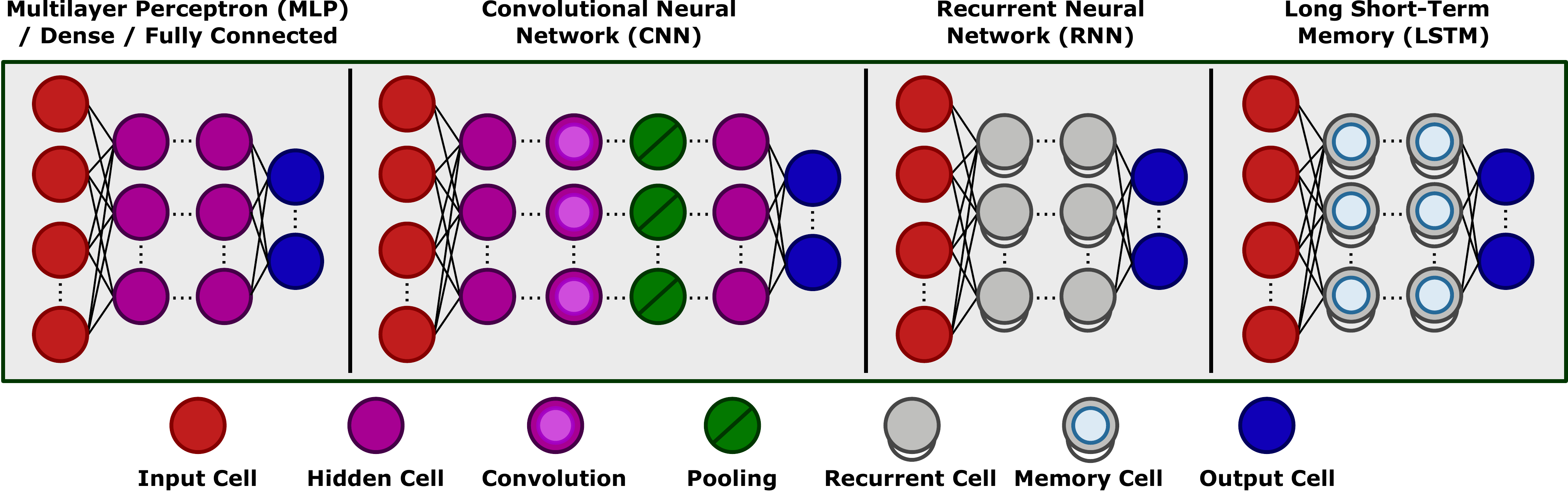}
		\caption{Popular \ac{ANN} structures. MLP/Dense/Fully Connected are typically well-suited for cross-sectional quantitative data, whereas \acp{RNN} and \acp{LSTM} networks are optimized for sequential data. \acp{CNN} are equipped for both types.}
		\label{fig:dnn}
	\end{figure*}     
	
    This tutorial on \ac{DL} accelerators within the biomedical sphere commences with a brief introduction to artificial and spiking neural networks. Next, we introduce the computational demands of \ac{DL} by shedding light on why they are power- and resource-intensive. This will justify the need for application specific hardware platforms. After that, we discuss recent hardware advances which have led to improvements in training and inference efficiency. These improvements ultimately guide us to viable edge inference engine options.
    
    After reviewing the literature on these \ac{DL} accelerators, we quantify the performance of various algorithms on different types of \ac{DL} processors. The results allow us to draw a perspective on the potential future of spike-based neuromorphic processors in the biomedical signal processing domain. Based on our analysis and perspective, we conjecture that, for edge processing, neuromorphic computing and \acp{SNN}~\cite{Chicca2014} will likely complement \ac{DL} inference engines, either through signaling anomalies in the data or acting as `intelligent always-on watchdogs' which continuously monitor the data being recorded, but only activate further processing stages if and when necessary.
    
    We expect this tutorial, review and perspective to provide guidance on the history and future of \ac{DL} accelerators, and the potential they hold for advancing healthcare. Our contributions are summarized as follows:
    \begin{itemize}
        \item Our paper is the first to discuss the use of three different emerging and established hardware technologies for facilitating \ac{DL} acceleration, with a focus on biomedical applications.
        \item  We provide tutorial sections on how one may implement a typical biomedical task on \acp{FPGA} or simulate it for deployment on memristive crossbars.
        \item Our paper is the first to discuss how event-based neuromorphic processors can complement \ac{DL} accelerators for biomedical signal processing. 
        \item We provide open-source code and data to enable the reproduction of our results.
    \end{itemize}
	
	The remainder of the paper is organized as follows. In Section~\ref{sec:nn}, we define the technical terminology that is used throughout this paper and cover the working principles of artificial and spiking neural networks. We also introduce a biomedical signal processing task for hand-gesture classification, which is used for benchmarking the different technologies and algorithms discussed in this paper. In Section~\ref{sec:dnn}, we step through the design, simulation, and implementation of \acp{DNN} using different hardware technologies. We show sample cases of how they have been deployed in healthcare settings. Furthermore, we demonstrate the steps and techniques required to simulate and implement hardware for the benchmark hand-gesture classification task using memristive crossbars and \acp{FPGA}. 
	
	In Section~\ref{sec:perspective}, we provide our perspective on the challenges and opportunities of both \acp{DNN} and \acp{SNN} for biomedical applications and shed light on the future of spiking neuromorphic hardware technologies in the biomedical domain. Section~\ref{sec:conclusion} concludes the tutorial.
	
	\section{Deep Artificial and Spiking Neural Networks}\label{sec:nn}
	
	\subsection{Nomenclature of Neural Network Architectures}
	
	Although most \acp{DNN} reported in literature are \acp{ANN}, \acp{DNN}  refer to more than one hidden layer, independently of whether the architecture is fully connected, convolutional, recurrent, \ac{ANN} or \ac{SNN}, or of any other structure. For example, the most widely used \ac{DNN} type in image processing, i.e. a \ac{CNN}, can be physically implemented as an \ac{ANN} or \ac{SNN}, and in both cases it would be `deep'. However, in this paper, whenever we use the terms `deep', \ac{DL}, or deep network, we refer to Deep Artificial Neural Networks. For Deep Spiking Neural Networks, we simply use the term \ac{SNN}.
	
	\subsection{Deep Artificial Neural Networks}
	Traditional \acp{ANN} and their learning strategies that were first developed several decades ago~\cite{Rumelhart1986} have, in the past several years, demonstrated unprecedented performance in a plethora of challenging tasks which are typically associated with human cognition. These have been applied to medical image diagnosis~\cite{Hollon2020} and medical text processing~\cite{Shickel2017}, using \acp{DNN}.
	
	Fig.~\ref{fig:dnn} illustrates a simplified overview of the structure of some of the most widely-used \acp{DNN}. The most conventional form of these architectures is the \ac{MLP}. Increasing the number of hidden layers of perceptron cells is widely regarded to improve hierarchical feature extraction which is exploited in various biomedical tasks, such as seizure detection from \ac{EEG}~\cite{Sayeed2019,Yang2020seizure}. \acp{CNN} introduce convolutional layers, which use spatial filters to encourage spatial invariance. \acp{CNN} often include pooling layers to downsample their outputs to reduce the search space for subsequent convolutional layers. \acp{CNN} have been widely used in medical and healthcare applications, as they are very well-suited for spatially structured data. Their use in medical image analysis~\cite{Tajbakhsh2016} will form a major part of our discussions in subsequent sections.
	
	\acp{RNN} are another powerful network architecture recently used both individually~\cite{Gao2019}, and in combination with \acp{CNN}~\cite{Zhang2018}, in biomedical applications. \acp{RNN} introduce recurrent cells with a feedback loop, and are especially useful for processing sequential data such as temporal signals and time-series data, e.g. \ac{ECG}~\cite{Zhang2018}, and medical text~\cite{Zhou2020}. The feedback loop in recurrent cells gives them a memory of previous steps and builds a dynamic awareness of changes in the input. The most well-known type of \acp{RNN} are \acp{LSTM} which are designed to mine patterns in data sequences using their short-term memory of distant events stored in their memory cells. \acp{LSTM} have been widely used for processing biomedical signals such as \acp{ECG}~\cite{Gao2019,Laitala2020}. Although there are many other variants of \ac{DNN} architectures, we will focus on these most commonly used types.        
	
	\subsubsection{Automatic hierarchical feature extraction} The above mentioned \acp{DNN} learn intricate features in data through multiple computational layers across various levels of abstraction~\cite{Goodfellow2016}. The fundamental advantage of \acp{DNN} is that they mine the input data features automatically, without the need for human knowledge in their supervised learning loop. This allows deep networks to learn complex features by combining a hierarchy of simpler features learned in their hidden layers~\cite{Goodfellow2016}. 
		
	\subsubsection{Learning algorithms} Learning features from data in a DNN, e.g. the networks shown in Fig.~\ref{fig:dnn}, is typically achieved by minimizing a loss function. In most cases, this is equivalent to finding the maximum likelihood using the cross-entropy between training data and the learned model distribution. Loss function minimization is achieved by optimizing the network parameters (weights and biases). This optimization process minimizes the loss function from the final network layer backward through all the network layers and is therefore called backpropagation. Widely used optimization algorithms in DNNs include Stochastic Gradient Descent (SGD) and those that use adaptive learning rates~\cite{Goodfellow2016}.
	
	\begin{table}[!b] 
    	\centering
    	\caption{Number of weights and \ac{MAC} operations in various CNN architectures for a single image and for video processing at 25 frames per second.} \label{tab:MAC}
        \begin{tabu} to 0.5\textwidth {p{0.175\textwidth}XXX}
    		\toprule
    		\textbf{Network architecture} & \textbf{Weights} & \textbf{MACs} & \textbf{$@$ 25 FPS} \\
    		\midrule
    		AlexNet & 61 M & 725 M & 18 B \\
    		ResNet-18 & 11 M & 1.8 B & 45 B \\
    		ResNet-50 & 23 M & 3.5 B & 88 B \\
    		VGG-19 & 144 M & 22 B & 550 B \\
    		OpenPose & 46 M & 180 B & 4500 B \\
    		MobileNet & 4.2 M & 529 M & 13 B \\ \bottomrule
    	\end{tabu}
    \end{table}
    
	\subsubsection{Backpropagation in DNNs is computationally expensive} Despite the continual improvement of hardware platforms for running and training \acp{DNN}, reducing their power consumption is a computationally formidable task. One of the dominant reasons is the feed-forward error backpropagation algorithm, which depends on thousands of epochs of computationally intensive \ac{VMM}  operations~\cite{Rumelhart1986}, using huge datasets that can exceed millions of data points. These operations, if performed on a conventional von Neumann architecture which has separate memory and processing units, will have a time and power complexity of order $\mathcal{O}(N^2)$ for multiplying a vector of length $N$ in a matrix of dimensions $N \times N$.  
	
	In addition, an artificial neuron in \acp{DNN} calculates a sum-of-products of its input-weight matrix pairs. For instance, a \ac{CNN} spatially structures the sum-of-products calculation into a \ac{VMM} operation. In digital logic, an adder tree can be used to accumulate a large number of values. This, however, becomes problematic in \acp{DNN} when one considers the sheer number of elements that must be summed together, as each addition requires one cycle.
	
	\subsubsection{Transfer learning} A major assumption when training \acp{DNN} is that both training and test samples are drawn from the same feature space and distribution. When the feature space and/or distribution changes, \acp{DNN} should be retrained. Rather than training a new model from scratch, trained parameters from an existing model can be fixed, tuned, or adapted~\cite{5288526}. This process of transfer learning can be used to greatly reduce the computational expense of training \acp{DNN}. 
	
    In the medical imaging domain, transfer learning from natural image datasets, particularly ImageNet~\cite{deng2009imagenet}, using standard large models and corresponding pretrained weights has become a de-facto method to speed up training convergence and to improve accuracy~\cite{Raghu2019TransfusionUT}. Transfer learning can also be used to leverage personalized anatomical knowledge accumulated over time to improve the accuracy of pre-trained \acp{CNN} for specific patients~\cite{elmahdy2020patient}, i.e., to perform patient-specific model tuning. This is an important topic in biomedical application domains, which will be further discussed in \ref{sec:patientspecific}.
	
	\subsection{DL Accelerators}
	In Table~\ref{tab:MAC}, we depict \color{black}some popular \ac{CNN} architectures, accompanied with the total number of weights, and \ac{MAC} operations that must be computed for a single image (input resolutions of 656$\times$468 for OpenPose, 224$\times$224 for the rest). This table highlights two key facts. Firstly, \acp{MAC} are the dominant operation of DNNs. Therefore, hardware implementations of \acp{DNN} should strive to parallelize a large number of \acp{MAC} to perform effectively. Secondly, there are many predetermined weights that must be called from memory. Reducing the energy and time consumed by reading weights from memory provides another opportunity to improve efficiency.
	
\begin{figure}[!t]
		\centering
		\includegraphics[width=0.47\textwidth]{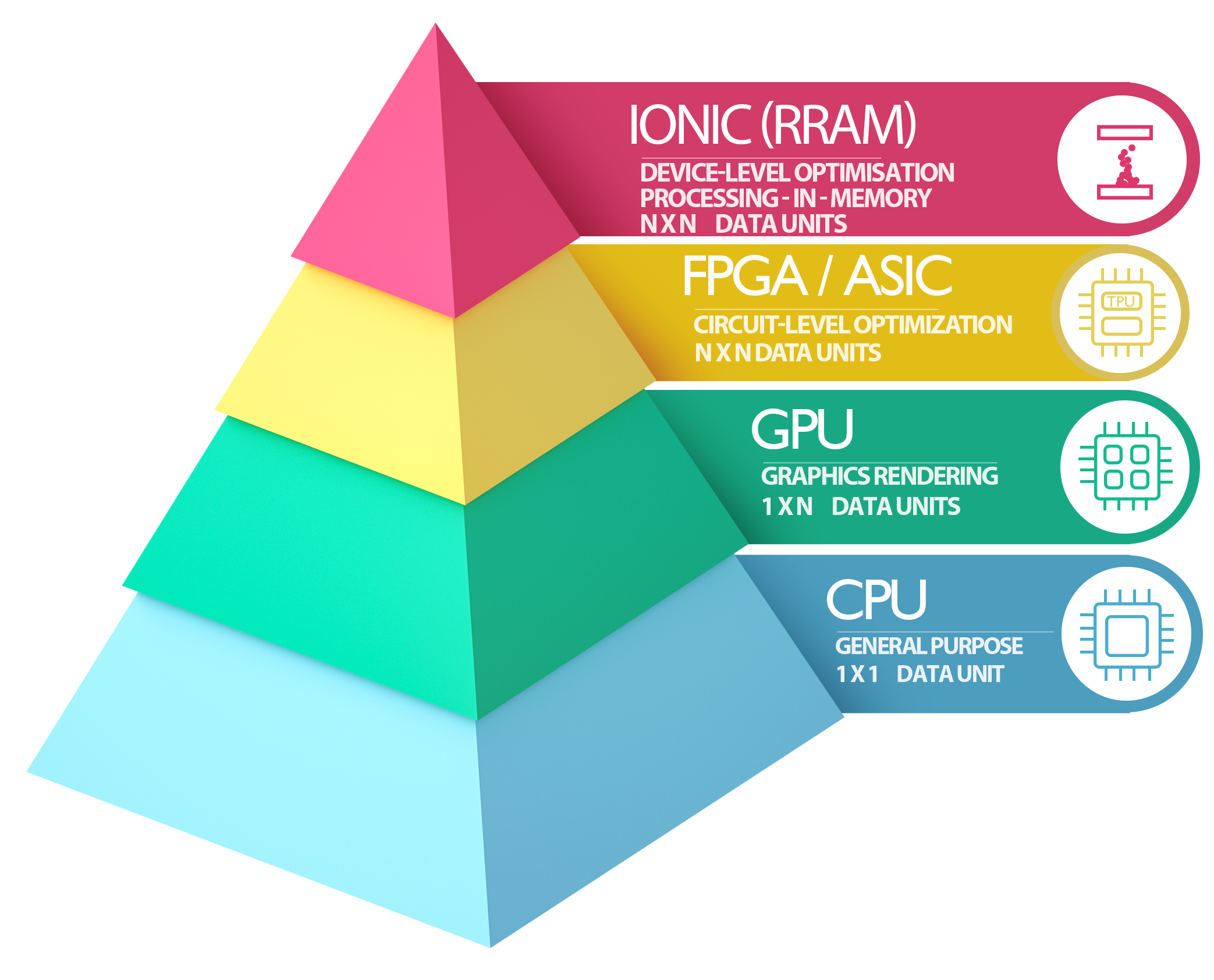}
		\caption{Typical hardware technologies for DNN acceleration. In this paper we cover the top two layers of the pyramid, which include specialized hardware technologies for high-performance training and inference of DNNs. While the apex is labelled RRAM, this is intended to broadly cover all programmable non-volatile resistive switching memories e.g. CBRAM, MRAM, PCM, etc.}
		\label{fig:pyramid}
	\end{figure}

	Consequently, significant research has been being conducted to achieve massive parallelism and to reduce memory access in \ac{DNN} accelerators, using different hardware technologies and platforms as depicted in Fig.~\ref{fig:pyramid}. Although these goals are towards general \ac{DL} applications, they can significantly facilitate fast and low-power smart \ac{PoC} devices~\cite{Xie2019} and healthcare \ac{IoT} systems. 
	
	In addition to conventional \ac{DL} accelerators, there have been significant research efforts to utilize biologically plausible \acp{SNN} for learning and cognition~\cite{Indiveri_Liu_15}. Spiking neuromorphic processors have also been used for biomedical signal processing~\cite{Corradi_etal15,Corradi_etal19,ceolini2020hand}. Below, we provide a brief introduction to \acp{SNN}, which will be discussed as a method complementary to \ac{DL} accelerators for efficient biomedical signal processing later in this paper. We will also perform comparisons among \acp{SNN} and \acp{DNN} in performing an \ac{EMG} processing task.   
	
	\subsection{Spiking Neural Networks}
\begin{figure*}[!t]
		\centering
		\includegraphics[width=1\textwidth]{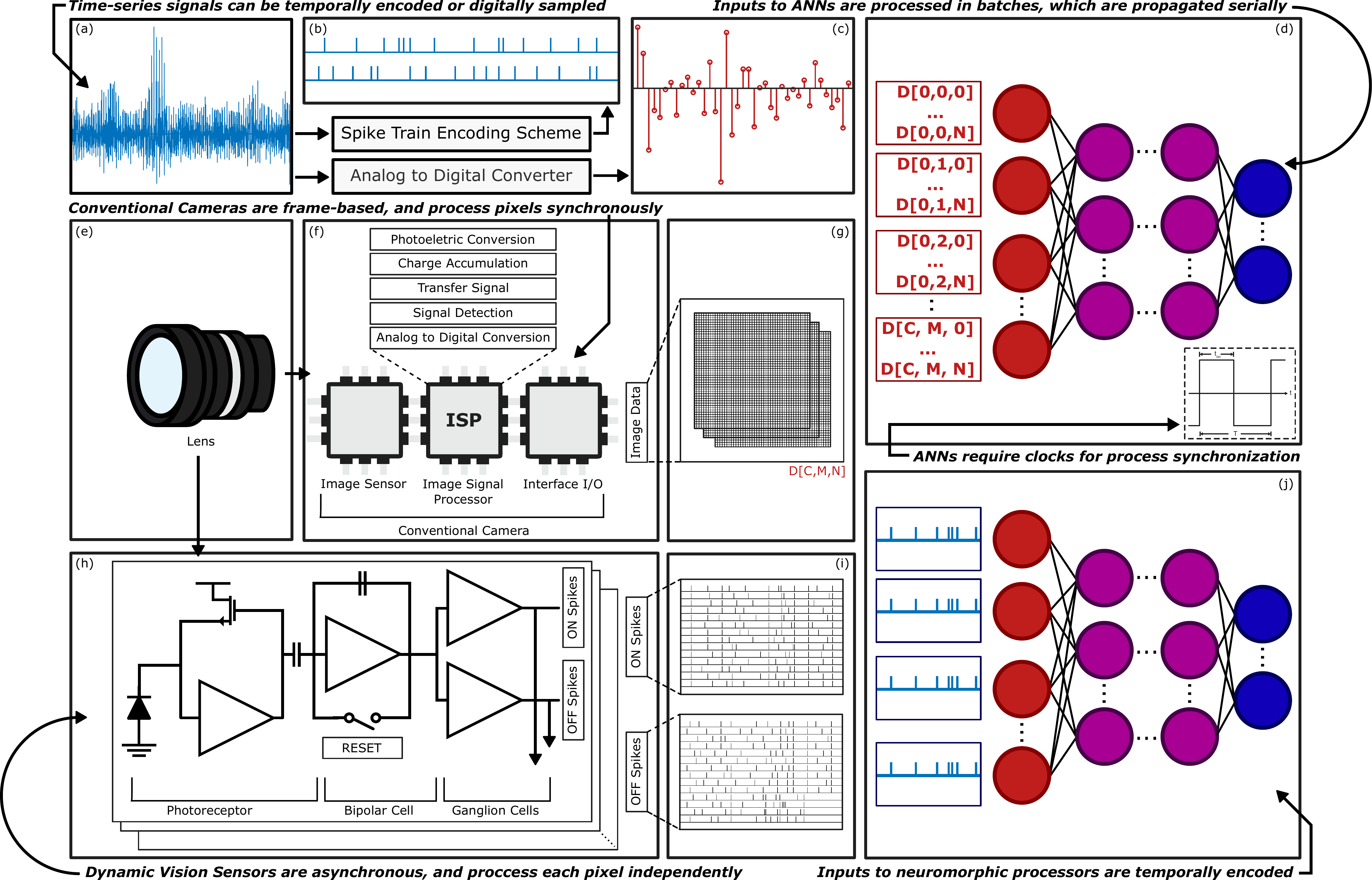}
		\caption{\acp{DNN} and \ac{SNN} neuromorphic processors adopt different operation models. In \acp{DNN}, inputs are processed in batches which propagate serially. Consequently, they require clocks for process synchronization. SNNs are asynchronous and process temporally encoded inputs independently. Time series signals, such as the EMG signal presented in (a) can be either (b) temporally encoded using spike train encoding schemes such as~\cite{Corradi_etal15}, before being fed into (j) neuromorphic processors, or (c) digitally sampled, before being concatenated into batches, to be fed into (d) \acp{DNN}. Similarly, photographs captured from (e) lenses can be (i) temporally encoded into spike trains using (h) \acp{DVS}~\cite{Lichtsteiner_etal2008} or (f) digitally encoded using conventional cameras to build (g) image frames.}
		\label{fig:ann_snn_comparison}
	\end{figure*}

    \acp{SNN} are neural networks that typically use Integrate-and-Fire neurons to dynamically process temporally varying signals (see Fig.~\ref{fig:ann_snn_comparison}(j)). By integrating multiple spikes over time, it is possible to reconstruct an analog value that represents the mean firing rate of the neuron. The mean firing rate is equivalent to the value of the activation function of \acp{ANN}. So in the mean firing rate limit, there is an equivalence between \acp{ANN} and \acp{SNN}. By using spikes as all-or-none digital events (Fig.~\ref{fig:ann_snn_comparison}(i)), \acp{SNN} enable the reliable transmission of signals across long distances in electronic systems. In addition, by introducing the temporal dimension, these networks can efficiently encode and process sequential data and temporally changing inputs~\cite{indiveri2019importance}. \acp{SNN} can be efficiently interfaced with event-based sensors since they only process events as they are generated. An example of such sensors is the \ac{DVS}, which is an event-based camera shown in Fig.~\ref{fig:ann_snn_comparison}(h). The \ac{DVS} consists of a logarithmic photo-detector stage followed by an operational transconductance amplifier with a capacitive-divider gain stage, and two comparators. The ON/OFF spikes are generated every time the difference between the current and previous value of the input exceeds a pre-defined threshold. The sign of the difference corresponds to the ON or OFF channel where the spike is produced. This is different to conventional cameras (Fig.~\ref{fig:ann_snn_comparison}(f)), which produce image frames (Fig.~\ref{fig:ann_snn_comparison}(g)). Intuitively, it makes sense to use asynchronous event-based sensor data in asynchronous \acp{SNN}, and synchronously generated frames (i.e., all pixels are given at a regular clock interval) through synchronous \acp{ANN}. But it is worth noting that conventional frames can be encoded as asynchronous spikes with frequencies that vary based on pixel intensity, and event streams can be integrated over time into synchronously generated time-surfaces \cite{eshraghian2018neuromorphic, lagorce2016hots}. Event-based sensors have been used to process biomedical signals~\cite{Corradi_etal15,Sharifshazileh_etal19} (Fig.~\ref{fig:ann_snn_comparison}(a)), which can be encoded to spike trains (Fig.~\ref{fig:ann_snn_comparison}(b)) to be processed by \acp{SNN} or be digitally sampled (Fig.~\ref{fig:ann_snn_comparison}(c)) for use in \acp{DNN} for learning and inference (Fig.~\ref{fig:ann_snn_comparison}(d)).   
    
	\subsection{Benchmarking on a Biomedical Signal Processing Task} \label{subsec:benchmark}
	In Section \ref{sec:dnn} we will present a use-case of bio-signal processing where \ac{FPGA} and memristive \ac{DNN} accelerators are implemented and simulated. These are later compared to equivalent existing implementations\footnote{\url{https://github.com/Enny1991/dvs_emg_fusion/blob/master/full_baseline.py}} using \ac{DNN} accelerators and neuromorphic processors from~\cite{ceolini2020hand}. 
	To perform comparisons, we use the same hand-gesture recognition task as in~\cite{ceolini2020hand}.

	Tasks such as prosthesis control can be performed using \ac{EMG} signals, hand-gesture classification, or a combination of both.  
	Here, the adopted hand-gesture dataset~\cite{ceolini2020hand} is a collection of 5 hand gestures recorded with two sensor modalities: muscle activity from a Myo armband that senses \ac{EMG} electrical activity in forearm muscles, and a visual input in the form of \ac{DVS} events. Moreover, the dataset provides accompanying video captured from a traditional frame-based camera, i.e., images from an \ac{APS} to feed \acp{DNN}. Recordings were collected from 21 subjects including 12 males and 9 females between the ages 25 and 35, and were taken over three separate sessions.
	
	For each implementation, we compare the mean and standard deviation of the accuracy obtained over a 3-fold cross validation, where each fold encapsulates all recordings from a given session. Additionally, for all implementations, we compare the energy and time required to perform inference on a single input, as well as the Energy-Delay Product (EDP), which is the average energy consumption multiplied by the average inference time.
	
	\section{DNN Accelerators towards healthcare and biomedical Applications} \label{sec:dnn}
	In this Section, we cover the use of CMOS and memristors in \ac{DL} acceleration. We discuss how they use different strategies to achieve two of the key DNN acceleration goals, namely \ac{MAC} parallelism and reduced memory access. We also discuss and review \acp{FPGA} as an alternative reconfigurable DNN accelerator platform, which has shown great promise in the healthcare and biomedical domains. 
	
	\subsection{CMOS DNN Accelerators} \label{subsec:cmosdnn}
	 General edge-AI CMOS accelerator chips can be used for DNN-enabled healthcare IoT and \ac{PoC} systems. Therefore, within this subsection, we first review a number of these chips and provide examples of potential healthcare applications they can accelerate. We will also explore some common approaches to CMOS-driven acceleration of \ac{AI} algorithms using massive \ac{MAC} parallelism and reduced memory access, which are useful for both edge-AI devices and offline data center scale acceleration. 
	 
    \subsubsection{Edge-AI DNN accelerators suitable for biomedical applications} 
	The research and market for \acp{ASIC}, which focus on a new generation of microprocessor chips dedicated entirely to machine learning and \acp{DNN}, have rapidly expanded in recent years. 
    Table~\ref{tab:cmosdnns} shows a number of these CMOS-driven chips, which are intended for portable applications. There are many other examples of AI accelerator chips (for a comprehensive survey see~\cite{Reuther2019}), but here we picked several prolific examples, which are designed specifically for \ac{DL} using \acp{DNN}, \acp{RNN}, or both. We have also included a few general purpose \ac{AI} accelerators from Google~\cite{Google2019}, Intel~\cite{Intel2019}, and Huawei~\cite{Huawei2019}.
    
    Although developed for general \acp{DNN}, the accelerators shown in Table~\ref{tab:cmosdnns} can efficiently realize portable smart \ac{DL}-based healthcare IoT and \ac{PoC} systems for processing image-based (medical imaging) or dynamic sequential medical data types (such as EEG and ECG). For instance, the table shows a few exemplar healthcare and biomedical applications that are picked based on the demonstrated capacity of these accelerators to run (or train~\cite{Lee2019}) various well-known CNN architectures such as VGG, ResNet, MobileNet, AlexNet, Inception, or \acp{RNN} such as LSTMs, or combined CNN-RNNs. It is worth noting that most of the available accelerators are intended for CNN inference, while only some~\cite{Shin2018,Yin2017,Lee2018} also include recurrent connections for \ac{RNN} acceleration. 
    
    The Table shows that the total power per chip in most of these devices is typically in the range of hundreds of mW, with a few exceptions consuming excessive power of around 10 Watts~\cite{Intel2019,Huawei2019}. This is required to avoid large heat sinks and to satisfy portable battery constraints. The Table also shows the computing capability per unit time (column `Computational Power (GOP/s)'). Regardless of power consumption, this column reveals the computational performance and consequently the size of a network one can compute per unit time. It is demonstrated that several of these chips can run large and deep CNNs such as VGG and ResNet, which enable them to perform complex processing tasks within a constrained edge power budget.

\begin{table*}[!t]
\centering
\caption{A number of recent edge-AI CMOS chips suitable for portable healthcare and biomedical applications.}
\label{tab:cmosdnns}
\begin{tabu} to \textwidth {p{0.15\textwidth}p{0.07\textwidth}p{0.07\textwidth}Xp{0.06\textwidth}Xp{0.275\textwidth}}
\toprule
\textbf{CMOS Chip} & \begin{tabular}[c]{@{}l@{}}\textbf{Core size}\\\textbf{(mm$^2$)}\\ \end{tabular} & \begin{tabular}[c]{@{}l@{}}\textbf{Technology}\\\textbf{(nm)~~}\end{tabular} & \begin{tabular}[c]{@{}l@{}}\textbf{Computational}\\\textbf{Power (GOP/s)}\\ \end{tabular} & \begin{tabular}[c]{@{}l@{}}\textbf{Power}\\\textbf{(mW)}\\ \end{tabular} & \begin{tabular}[c]{@{}l@{}}\textbf{Power Efficiency}\\\textbf{(TOPS/W)}\\ \end{tabular} & \begin{tabular}[c]{@{}l@{}}\textbf{Potential Mobile and Edge-based Health-}\\\textbf{care and Medical Applications}\\ \end{tabular} \\
\midrule
Cambricon-x~\cite{Zhang2016} & 6.38 & 65 & 544 & 954 & 0.5 & \ac{ECog} analysis using a sparse VGG~\cite{Zhang2017} for \ac{PoC} diagnosis of cardiovascular diseases \\ 
\midrule
Eyeriss~\cite{Chen2016} & 12.25 & 65 & 17-42 & 278 & 0.06--0.15 & - Mobile Image-based cancer diagnosis using VGG-16~\cite{Guan2019}, \\
 &  &  &  &  &  & - Mobile diagnosis tool based on AlexNet for radiology, cardiology gastroenterology imaging~\cite{Tajbakhsh2016} \\ 
\midrule
Origami~\cite{Cavigelli2016} & 3.09 & 65 & 196 & 654 & 0.8 & - Smart healthcare \ac{IoT} edge device for heart health monitoring using a \ac{CNN}-based \ac{ECG} analysis~\cite{Azimi2018} \\
 &  &  &  &  &  & - Two-stage end-to-end \ac{CNN} for human activity recognition~\cite{8684824}\\
\midrule
ConvNet processor~\cite{Moons2016} & 2.4 & 40 & 102 & 25-287 & 0.3--2.7 & \ac{PoC} Ultrasound processing using AlexNet~\cite{Blaivas2019} \\ 
\midrule
Envision~\cite{Moons2017} & 1.87 & 28 & 76-408 & 7.5-300 & 0.8--10 & Multi-layer \ac{CNN} for \ac{EEG}/\ac{ECog} feature extraction for epileptogenicity for epilepsy diagnosis on edge~\cite{Hosseini2017} \\ 
\midrule
Neural processor~\cite{Song2019} & 5.5 & 8 & 1900-7000 & 39--1500 & 4.5-11.5 & On edge classification of skin cancer using Inception V3 \ac{CNN}~\cite{Esteva2017} \\ 
\midrule
LNPU~\cite{Lee2019} & 16 & 65 & 600 & 43-367 & 25 & - On edge learning/inference using VGG-16 for cancer diagnosis~\cite{Guan2019}, \\
 &  &  &  &  &  & -~On edge AlexNet learning/inference for radiology, cardiology, gastroenterology imaging diagnosis~\cite{Tajbakhsh2016} \\ 
\midrule
DNPU~\cite{Shin2018} & 16 & 65 & 300-1200 & 35-279 & 2.1--8.1 & Parallel and Cascade \ac{RNN} and \ac{CNN} for ac{ECG} analysis for \ac{BCI}~\cite{Zhang2018} \\ 
\midrule
Thinker~\cite{Yin2017} & 14.44 & 65 & 371 & 293 & 1--5 & - \ac{PoC} conversion of respiratory organ motion ultrasound into MRI using a long-term recurrent \ac{CNN}~\cite{Preiswerk2018} \\ 
\midrule
UNPU~\cite{Lee2018} & 16 & 65 & 346-7372 & 3.2-297 & 3.08--50.6 & -~Intelligent pre-diagnosis medical support/consultation using a \ac{CNN}-\ac{RNN}~\cite{Zhou2020} \\
 &  &  &  &  &  & -~A CNN-RNN for respiratory sound
classification in wearable devices enabled by patient specific model tuning~\cite{9040275} \\  &  &  &  &  &  & -~A CNN-LSTM for missing Photoplethysmographic data prediction~\cite{roy2020device} \\
\midrule
Google Edge TPU~\cite{Google2019} & 25 & - & 4000 & 2000 & 2 & - Low-cost and easy-to-access skin cancer detection using MobileNet V1 \ac{CNN}~\cite{Sahu2018} \\
 &  &  &  &  &  & - On edge health monitoring for fall detection using \ac{LSTM}s~\cite{Queralta2019} \\
 &  &  &  &  &  & - Robust long-term decoding in intracortical \acp{BMI} using MLP and \ac{ELM} networks~\cite{8944156}\\
\midrule
Intel Nervana NNP-I 1000 (Spring Hill)~\cite{Intel2019} & - & 10 & 48000 & 10000 & 4.8 & - Diagnosis using chest X-ray classification on ResNet \ac{CNN} family~\cite{Baltruschat2019} \\
 &  &  &  &  &  & - Long term bowel sound segmentation using a \ac{CNN}~\cite{zhao2020long}\\
\midrule
Huawei Ascend 310 \cite{Huawei2019} & - & 12 & 16000 & 8000 & 2 & - Cardiovascular monitoring for arrhythmia diagnosis from \ac{ECG} using an \ac{LSTM}~\cite{Gao2019}, \\
 &  &  &  &  &  & - Health monitoring by heart rate variability analysis using \ac{ECG} analysis by a bidirectional \ac{LSTM}~\cite{Laitala2020} \\
\bottomrule
\end{tabu}
\end{table*}

    For instance, it has been previously shown in~\cite{Zhang2017} that VGG CNN (shown to be compatible with Cambricon-x \cite{Zhang2016}), can successfully analyze \ac{ECog} signals. Therefore, considering the power efficiency of Cambricon-x, it can be used to implement a portable automatic \ac{ECog} analyzer for \ac{PoC} diagnosis of various cardiovascular diseases~\cite{Zamzmi2020}. Similarly, Eyeriss~\cite{Chen2016} can run VGG-16, which is shown to be effective in diagnosing thyroid cancer~\cite{Guan2019}. In addition, Eyeriss can run AlexNet for several different medical imaging applications~\cite{Tajbakhsh2016}. Therefore, Eyeriss can be used as a mobile diagnostic tool that can be integrated into or complement medical imaging systems at the \ac{PoC}. Origami~\cite{Cavigelli2016} is another CNN accelerator chip, which can be used for other healthcare applications based on a CNN. For instance, \cite{Azimi2018} proposes a CNN-based ECG analysis for heart monitoring, or~\cite{8684824} introduces a two-stage end-to-end \ac{CNN} for human activity recognition for elderly and rehailitation monitoring, whereas Origami can be used to develop a smart healthcare IoT edge device. Similarly, the CNN processor proposed in~\cite{Moons2016} is shown to be able to run AlexNet, which can be deployed in a \ac{PoC} ultrasound image processing system~\cite{Blaivas2019}. Envision~\cite{Moons2017} is another accelerator that has the capability to run large-scale CNNs. It can also be used as an edge inference engine for a multi-layer CNN for EEG/\ac{ECog} feature extraction for epilepsy diagnosis~\cite{Hosseini2017}. Neural processor \cite{Song2019} is another CNN accelerator that is shown to be able to run Inception V3 CNN, which can be used for skin cancer detection~\cite{Esteva2017} at the edge. LNPU~\cite{Lee2019} is the only CNN accelerator shown in Table~\ref{tab:cmosdnns}, which unlike the others can perform both learning and inference of a deep network such as AlexNet and VGG-16, for applications including on edge medical imaging~\cite{Tajbakhsh2016} and cancer diagnosis~\cite{Guan2019}.
    
    Unlike the above discussed chips that are capable of running only CNNs, DNPU \cite{Shin2018}, Thinker~\cite{Yin2017}, and UNPU~\cite{Lee2018} are capable of accelerating both CNNs and RNNs. This feature makes them suitable for a wider variety of edge-based biomedical applications such as \ac{ECG} analysis for \ac{BCI} using a cascaded RNN-CNN~\cite{Zhang2018}, \ac{PoC} MRI construction from motion ultrasounds using a long-term recurrent CNN~\cite{Preiswerk2018}, intelligent medical consultation using a CNN-RNN~\cite{Zhou2020}, respiratory sound classification in wearable devices enabled by patient specific model tuning using a CNN-RNN~\cite{9040275}, or on-chip online and personalized prediction of missing Photoplethysmographic data~\cite{roy2020device}.
    
     \begin{figure*}[!t]
		\centering
		\includegraphics[width=1\textwidth]{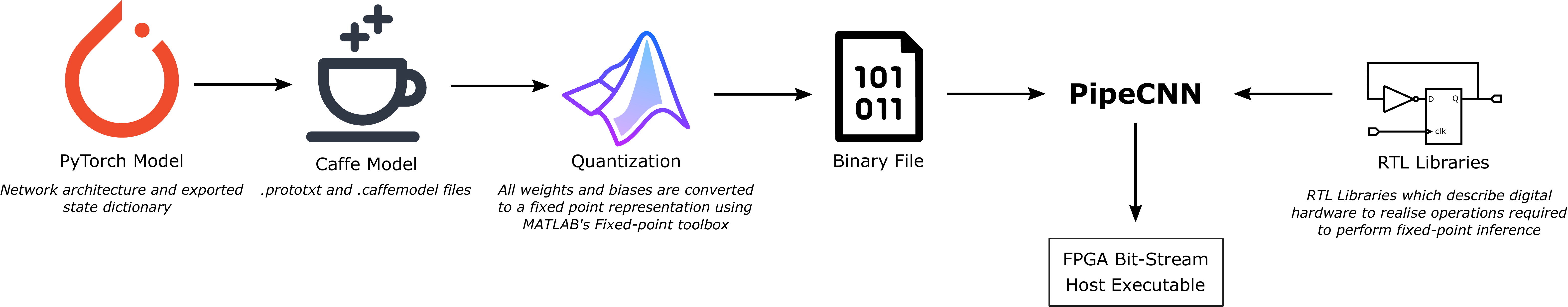}
		\caption{Compilation flow used to deploy an EMG classification CNN to an OpenVINO FPGA adopting fixed-point number representations using OpenCL.}
		\label{fig:flow}
	\end{figure*}
	
    Table~\ref{tab:cmosdnns} lists three general purpose \ac{AI} accelerator chips, which have been deployed for low-cost and easy-to-access skin cancer detection using MobileNet V1 CNN~\cite{Sahu2018}, on edge health monitoring for fall detection using LSTMs~\cite{Queralta2019}, chest X-ray analysis using ResNet CNN~\cite{Baltruschat2019}, long term bowel sound monitoring and segmentation using a CNN~\cite{zhao2020long}, cardiovascular arrhythmia detection from ECG using an LSTM~\cite{Gao2019}, or heart rate variability analysis from ECG signals through a bidirectional LSTM~\cite{Laitala2020}, just to name a few. These general-purpose chips have the potential to be used for other biomedical edge-based applications such as robust long-term decoding in intracortical \acp{BMI} using MLP and \ac{ELM} networks in a sparse ensemble machine learning platform~\cite{8944156}.
    
    In addition to the edge-AI CNN and RNN acceleration chips or general ML chips mentioned in Table~\ref{tab:cmosdnns}, there have been other works that have developed custom CMOS platforms for biomedical applications. Examples of these CMOS designs include~\cite{7348721} that has developed a 128-Channel \ac{ELM}-based neural decoder for \ac{BMI}, and \cite{8490222} that has implemented an autoencoder neural network as part of a neural interface processor for brain-state classification and programmable-waveform neurostimulation.    
    
    \subsubsection{Common approaches to CMOS-driven DL acceleration}
	Accelerators will typically target either data center use or embedded `edge-AI' acceleration. Edge chips, such as those discussed above, must operate under restrictive power budgets (e.g., within thermal limits of 5~W) to cope with portable battery constraints. While the scale of tasks, input dimension capacity, and clock speeds will differ between edge-AI and modular data center racks, both will adopt similar principles in the tasks they seek to optimize.
	
	Most of the accelerator chips, such as those discussed in Table~\ref{tab:cmosdnns}, use similar optimization strategies involving reduced precision arithmetic \cite{Lee2019,Lee2018,Moons2016,Moons2017} to improve computational throughput. This is typically combined with architectural-level enhancements \cite{Shin2018,Yin2017,Zhang2016,Chen2016,Song2019} to either reduce data movement (using in- or near-memory computing), heightened parallelism, or both. In addition, there are many other approaches commonly used to make neural network implementations more efficient. Examples of these include tensor decomposition, pruning, and mixed-precision data representation, which are often integrated in hardware with in-memory and near-memory computing. A thorough review of these approaches can be found in~\cite{deng2020model} and~\cite{sze2017efficient}.
	
	Sequential and combinational logic research is largely matured, so outside of emerging memory technologies, the dominant hardware benefits are brought on by optimizing data flow and architecture. An early example is the neuFlow system-on-chip (SoC) processor which relies on a grid of processing tiles, each made up of a bank of processing operators and a multiplexer based on-chip router \cite{pham2012neuflow}. The processing operator can serially perform primitive computation (MUL, DIV, ADD, SUB, MAX), or a parallelized 1D/2D convolution. The router configures data movement between tiles to support streaming data flow graphs.
	
	Since the development of neuFlow, over 100 startups and companies have developed, or are developing, machine learning accelerators. The Neural Processing Unit (NPU) \cite{putnam2014reconfigurable} generalizes the work from neuFlow by employing eight processing engines which each compute a neuron response: multiplication, accumulation, and activation. If a program could be partitioned such that a segment of it can be calculated using \acp{MAC}, then it would be partially computed on the NPU. This made it possible to go beyond \ac{MLP} neural networks. The NPU was demonstrated to perform Sobel edge detection and fast Fourier transforms as well. 
	
    NVIDIA coupled their expertise in developing \acp{GPU} with machine learning dedicated cores, namely, tensor cores, which are aimed at demonstrating superior performance over regular Compute Unified Device Architecture (CUDA) cores \cite{jia2018dissecting}. Tensor cores target mixed-precision computing, with their NVIDIA Tesla V100 \ac{GPU} combining 672 tensor cores on a single unit. By merging the parallelism of \acp{GPU} with the application specific nature of tensor cores, their \acp{GPU} are capable of energy efficient general compute workloads, as well as 12 \acp{TFLOPS} of matrix arithmetic. 
    
    Although plenty of other notable architectures exist (see Table~\ref{tab:cmosdnns}), a pattern begins to emerge, as most specialized processors rely on a series of sub-processing elements which each contribute to increasing throughput of a larger processor \cite{sze2017efficient, deng2020model}. Whilst there are plenty of ways to achieve \ac{MAC} parallelism, one of the most renowned techniques is the systolic array, and is utilized by Groq \cite{abtsthink} and Google, amongst numerous other chip developers. This is not a new concept: systolic architectures were first proposed back in the late 1970s~\cite{Kung1979,Kung1982}, and have become widely popularized since powering the hardware DeepMind used for the AlphaGo system to defeat Lee Sedol, the world champion of the board game Go in October 2015. Google also uses systolic arrays to accelerate \acp{MAC} in their \acp{TPU}, just one of many CMOS \acp{ASIC} used in DNN processing \cite{Jouppi2017}.
	
	\subsection{FPGA DNNs}
	\acp{FPGA} are fairly low-cost reconfigurable hardware that can be used in almost any hardware prototyping and implementation task, significantly shortening the time-to-market of an electronic product. They also provide parallel computation, which is essential when simultaneous data processing is required such as processing multiple \ac{ECG} channels in parallel. Furthermore, there exists a variety of High Level Synthesis (HLS) tools and techniques~\cite{Stone2010,Guo2019} that facilitate FPGA prototyping without the need to directly develop time-consuming low-level Hardware Description Language (HDL) codes~\cite{Cong2011}. These tools allow engineers to describe their targeted hardware in high-level programming languages such as C to synthesize them to Register Transfer Level (RTL). The tools then offload the computational-critical RTL to run as kernels on parallel processing platforms such as FPGAs~\cite{Lammie2019accelerating}. 
	
	\subsubsection{Accelerating DNNs on FPGAs}
	FPGAs have been previously used to realize mostly inference~\cite{Guo2019,Lammie2019a,Carreras2020}, and in some cases training of DNNs with reduced-precision-data~\cite{Lammie2020a}, or hardware-friendly approaches~\cite{Lammie2019b}. For a comprehensive review of previous FPGA-based DNN accelerators, we refer the reader to~\cite{Guo2019}.
	
	Here, we demonstrate an example of accelerating \acp{DNN} to benchmark the biomedical signal processing task explained in subsection~\ref{subsec:benchmark}. For our acceleration, we use fixed-point parameter representations on a Starter Platform for OpenVINO Toolkit FPGA using OpenCL. OpenCL~\cite{Stone2010} is an HLS framework for writing programs that execute across heterogeneous platforms. OpenCL specifies programming languages (based on C99 and C++11) for programming the compute devices and Application Programming Interfaces (APIs) to control and execute its developed kernels on the devices, where depending on the available computation resources, an accelerator can pipeline and execute all work items in parallel or sequentially.
	
	Fig.~\ref{fig:flow} depicts the compilation flow we adopted. The trained \ac{DNN} PyTorch model is first converted to \textit{.prototxt} and \textit{.caffemodel} files using Caffe. All weights and biases are then converted to a fixed point representation using MATLAB's Fixed-point toolbox using word length and fractional bit lengths defined in~\cite{Wang2017a}, prior to being exported as a single binary \textit{.dat} file for integration with PipeCNN, which is used to generate the necessary RTL libraries, and to perform compilation of the host executable and the FPGA bit-stream. We used Intel's FPGA SDK for OpenCL 19.1, and provide all files used during the compilation shown in Fig.~\ref{fig:flow} in a publicly accessible complementary GitHub repository\footnote{\url{https://github.com/coreylammie/TBCAS-Towards-Healthcare-and-Biomedical-Applications/blob/master/FPGA/}}.
		
	\subsubsection{FPGA-based DNNs for biomedical applications}
	Despite the many FPGA-based \ac{DNN} accelerators available~\cite{Guo2019}, only a few have been developed specifically for biomedical applications such as ECG anomaly detection~\cite{Wess2017}, or real-time mass-spectrometry data analysis for cancer detection~\cite{Sanaullah2018}, where the authors show that application-specific parameter quantization and customized network design can result in significant inference speed-up compared to both CPU and GPU. In addition, the authors in~\cite{Shrivastwa2018} have developed an FPGA-based \ac{BCI}, in which a \ac{MLP} is used for reconstructing \ac{ECog} signals. In~\cite{Chen2018}, the authors have implemented an \ac{EEG} processing and neurofeedback prototype on a low-power but low-cost FPGA and then scaled it on a high-end Ultra-scale Virtex-VU9P, which has achieved 215 and 8 times power efficiency compared to CPU and GPU, respectively. For the EEG processing, they developed an \ac{LSTM} inference engine.      
	
	It is projected that, by leveraging specific algorithmic design and hardware-software co-design techniques, \acp{FPGA} can provide \textgreater 10 times energy-delay efficiency compared to state-of-the-art \acp{GPU} for accelerating \ac{DL}~\cite{Guo2019}. This is significant for realizing portable and reliable healthcare applications. However, \ac{FPGA} design is not as straightforward as high-level designs conducted for \ac{DL} accelerators and requires skilled engineers and stronger tools, such as those offered by the \ac{GPU} manufacturers.

	\subsection{Memristive DNNs} \label{subsec:memdnn}
	To achieve the two aforementioned key DNN acceleration goals, i.e. massive \ac{MAC} parallelism and reduced memory access, many studies have leveraged memristors~\cite{Burr2015,Ambrogio2018,Yao2020,eshraghian2019analog} as weight elements in their \ac{DNN} and \ac{SNN}~\cite{Azghadi2017,rahimi2020complementary} architectures. Memristors are often referred to as the fourth fundamental circuit element, and can adapt their resistance (conductance) to changes in the applied current or voltage. This is similar to the adaptation of neural synapses to their surrounding activity while learning. This adaptation feature is integral to the brain's in-memory processing ability, which is missing in today's general purpose computers. This in-situ processing can be utilized to perform parallel \ac{MAC} operations inside memory, hence, significantly improving \ac{DNN} learning and inference. This is achieved by developing memristive crossbar neuromorphic architectures, which are projected to achieve approximately 2500-fold reduction in power and a 25-fold increase in acceleration, compared to state-of-the-art specialized hardware such as \acp{GPU}~\cite{Burr2015}.
	
    \begin{figure}[!b]
		\centering
		\includegraphics[width=0.47\textwidth]{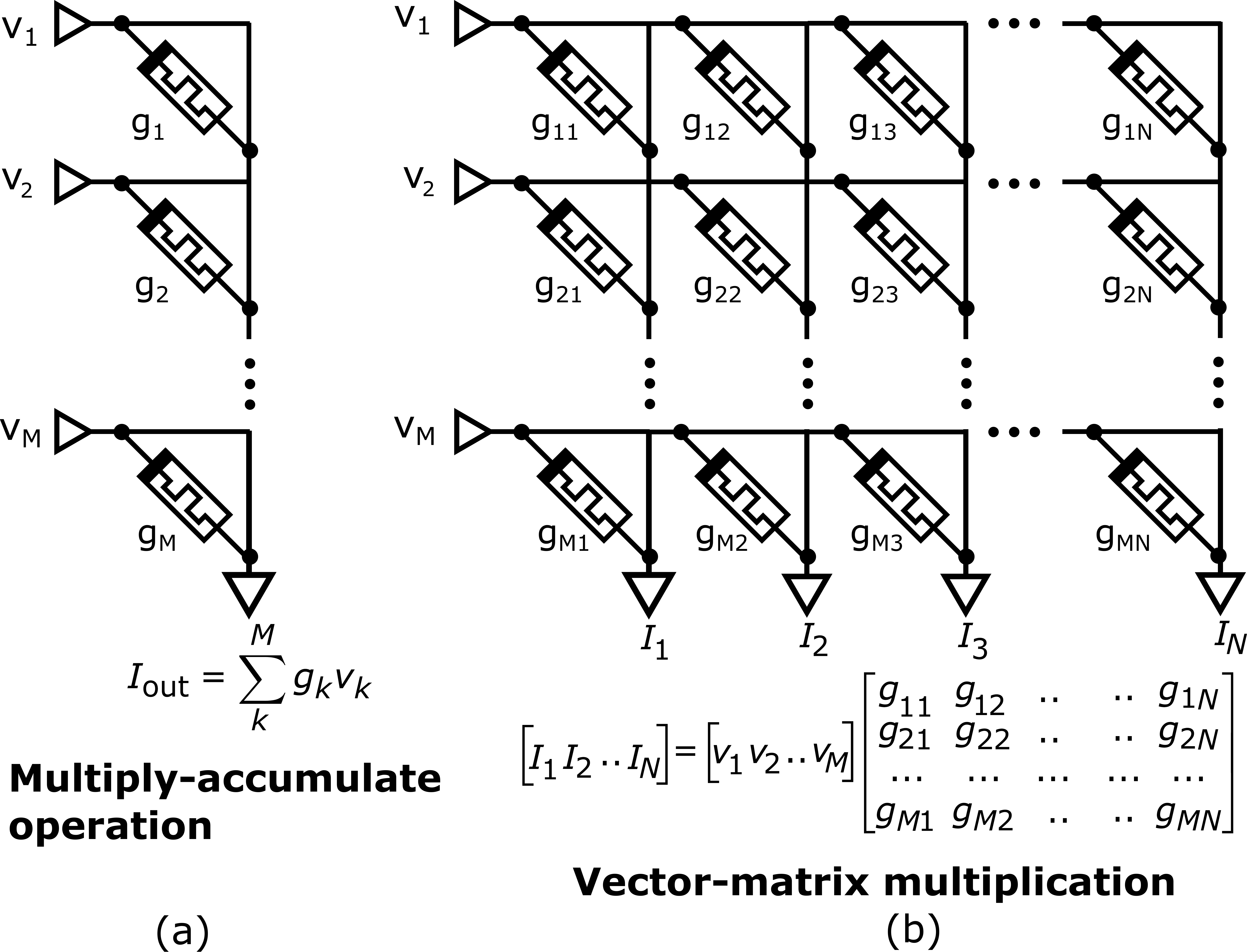}
		\caption{Memristive crossbars can parallelize (a) analog \ac{MAC} and (b) \ac{VMM} operations. Here, V represents the input vector, while conductances in the crossbar represent the matrix.}
		\label{fig:xbar}
	\end{figure}
	
	\subsubsection{Memristive crossbars for parallel MAC and VMM operations} A memristive crossbar that can be fabricated using a variety of device technologies~\cite{rahimi2020complementary,Xia2019} can perform analog \ac{MAC} operations in a single time-step (see Fig.~\ref{fig:xbar}(a)). This reduces the time complexity to its minimum ($\mathcal{O}(1)$), and is achieved by carrying out multiplication at the place of memory, in a non-von Neumann structure. Using this well-known approach, \ac{VMM} can be parallelized as demonstrated in Fig.~\ref{fig:xbar}(b), where the vector of size $M$ represents input voltage signals ($[V_1 .. V_M]$). These voltages are applied to the rows of the crossbar, while the matrix (of size $M \times N$), whose elements are represented as conductances (resistances), is stored in the memristive components at each cross point. 
	Taking advantage of the basic Ohm's law ($I=V.G$), the current summed in each crossbar column represents one element of the resulting multiplication vector of size $N$.	
	 
	\subsubsection{Mapping memristive crossbars to DNN layers} 
\begin{figure*}[!t]
		\centering
		\includegraphics[width=1\textwidth]{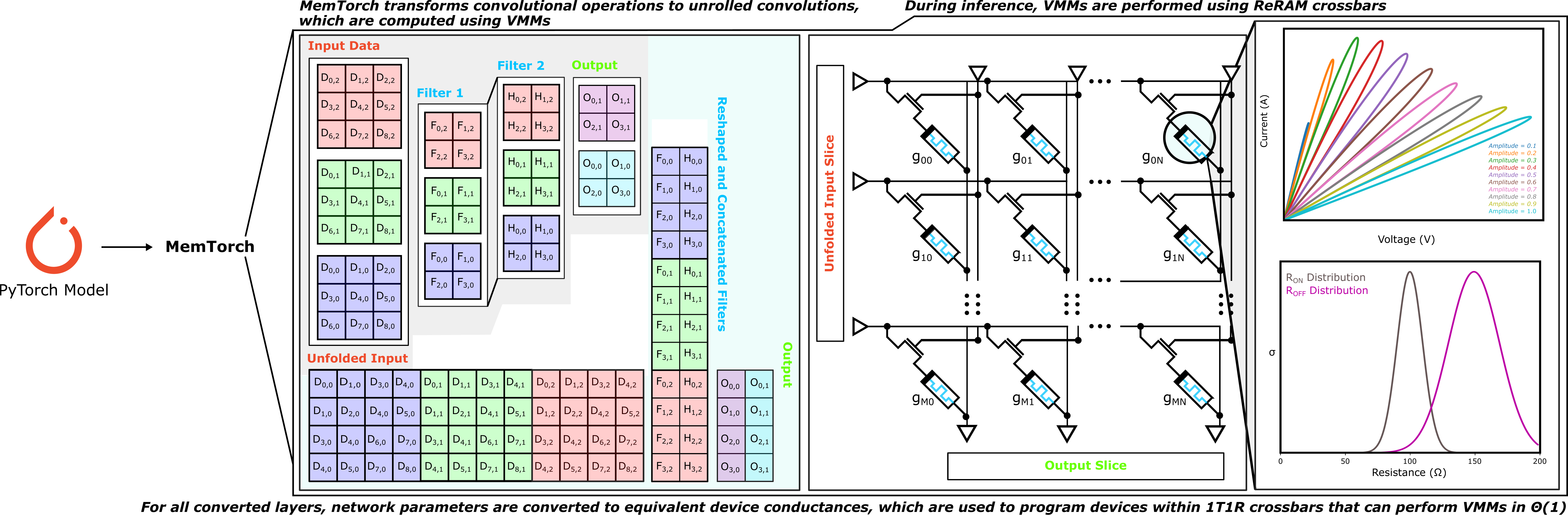}
		\caption{Conversion process of a DNN trained in PyTorch and mapped to a Memristive DNN using MemTorch~\cite{Lammie2020}, to parallelize MVMs using 1T1R memristive crossbars and to take into account memristor variability including finite number of conductance states and non-ideal $R_{\textnormal{ON}}$ and $R_{\textnormal{OFF}}$ distributions.}
		\label{fig:memtorch_flow}
	\end{figure*}

Although implementing fully-connected DNN layers is straightforward by mapping the weights to crossbar point memristors and having the inputs represented by input voltages, implementing a complex \ac{CNN} requires mapping techniques to convert convolution operations to \ac{MAC} operations. A popular approach to perform this conversion is to use an unrolling (unfolding) operation that transforms the convolution of input feature maps and convolutional filters to \ac{MAC} operations. We have developed a software platform named \emph{MemTorch}~\cite{Lammie2020}, that will be introduced in subsequent sections, to perform this mapping as well as a number of other operations, for converting \acp{DNN} to \acp{MDNN}. 
    The mapping process implemented in MemTorch is illustrated in the left panel in Fig.~\ref{fig:memtorch_flow}. The figure shows that the normal input feature maps and convolutional filters (shown in gray shaded area) are unfolded and reshaped (shown in the cyan shaded area) to be compatible with memristive crossbar parallel \ac{VMM} operations. It is worth noting that the convolutional filters that can be applied to the input feature maps have a direct relationship with the required crossbar sizes. Furthermore, the resulting hardware size depends on the size of the input feature maps~\cite{Lammie2019}.
    
	\subsubsection{Peripheral circuitry for memristive DNNs}
	In addition to the memristive devices that are used as programmable elements in \ac{MDNN} architectures, various peripheral circuitry is required to perform feed-forward error-backpropagation learning in \acp{MDNN}~\cite{Yao2020}. This extra circuitry may include: (i) a conversion circuit to translate the input feature maps to input voltages, which for programming memristive devices are usually \ac{PWM} circuits, (ii) current integrators or sense amplifiers, which pass the current read from every column of the memristive crossbar to (iii) analog to digital converters (ADCs), which pass the converted voltage to (iv) an activation function circuit, for forward propagation, and for backward error propagation (v) the activation function derivative circuit. Other circuits required in the error backpropagation path include (vi) backpropagation values to \ac{PWM} voltage generators, (vii) backpropagation current integrators, and (viii) backpropagation path ADCs. In addition, an update module that updates network weights based on an algorithm such as SGD is required, which is usually implemented in software. After the update, the new weight values should be written to the memristive crossbar, which itself requires Bit-Line (BL) and Word-line (WL) switch matrices to address the memristors for update, as well as a circuit to update the memristive weights. There are different approaches to implement this circuit such as that proposed in~\cite{Krestinskaya2018}, while others may use software ex-situ training where the new weight values are calculated in software and transferred to the physical memristors through peripheral circuitry~\cite{Yao2020}.

	\subsubsection{Memristive device nonidealities}
	Although ideal memristive crossbars have been projected to remarkably accelerate DNN learning and inference and drastically reduce their power consumption~\cite{Burr2015,Ambrogio2018}, device imperfections observed in experimentally fabricated memristors impose significant performance degradation when the crossbar sizes are scaled up for deployment in real-world DNN architectures, such as those required for healthcare and biomedical applications discussed in subsection~\ref{subsec:cmosdnn}. 
	These imperfections include nonlinear asymmetric and stochastic conductance (weight) update, device temporal and spatial variations, device yield, as well as limited on/off ratios~\cite{Burr2015}. To minimize the impact of these imperfections, specific peripheral circuitry and system-level mitigation techniques have been used~\cite{Yu2015}. However, these techniques add significant computation time and complexity to the system. It is, therefore, essential to take the effect of these nonidealities into consideration before utilizing memristive DNNs for any healthcare and medical applications, where accuracy is critical. In addition, there is a need for a unified tool that reliably simulates the conversion of a pre-trained \ac{DNN} to a \ac{MDNN}, while critically considering experimentally modeled device imperfections~\cite{Lammie2020}.         
	
	\subsubsection{Conversion of DNN to MDNN while considering memristor nonidealities}
	Due to the significant time and energy required to train new large versions of DNNs for challenging cognitive tasks, such as biomedical and healthcare data processing ~\cite{mckinney2020international,Bien2018}, the training of the algorithms is usually undertaken in data centers~\cite{mckinney2020international,Jouppi2017}. The pretrained DNN can then be transferred to be used on memristive crossbars. There are several different frameworks and tools that can be used to simulate and facilitate this transition~\cite{Ankit2019}. In a recent study, we have developed a comprehensive tool named MemTorch, which is an open source, general, high-level simulation platform that can fully integrate any behavioral or experimental memristive device model into crossbar architectures to design \acp{MDNN}~\cite{Lammie2020}. 
	
	Here, we utilize the benchmark biomedical signal processing task explained in subsection~\ref{subsec:benchmark} to demonstrate how pretrained \acp{DNN} can be converted to equivalent \acp{MDNN}, and how non-ideal memristive devices can be simulated within \acp{MDNN} prior to hardware realization. The conversion process, which can be generalized to other biomedical models using MemTorch, is depicted in Fig.~\ref{fig:memtorch_flow}. 
	
	 The targeted MDNNs are constructed by converting linear and convolutional layers from PyTorch pre-trained \acp{DNN} to memristive equivalent layers employing 1-Transistor-1-Resistor (1T1R) crossbars. A double-column scheme, in which two crossbars are used to represent positive and negative weight values, is used to represent network weights within memristive crossbars. The converted MDNN models are tuned using linear regression, as described in~\cite{Lammie2020}. The complete and detailed process and the source code of the network conversion for the experiments shown in this subsection are provided in a publicly accessible complementary Jupyter Notebook\footnote{\url{https://github.com/coreylammie/TBCAS-Towards-Healthcare-and-Biomedical-Applications/blob/master/MemTorch.ipynb}}.
	
	During the conversion, any memristor model can be used. For the benchmark task, a reference VTEAM model~\cite{Kvatinsky2015} is instantiated using parameters from Pt/Hf/Ti \ac{RRAM} devices~\cite{6069527}, to model all memristive devices within converted linear and convolutional layers. As already mentioned, memristive devices have inevitable variability, which should be taken into account when implementing an \acp{MDNN} for learning and/or inference. Also, depicted in Fig.~\ref{fig:memtorch_flow} are visualizations of two non-ideal device characteristics: the finite number of conductance states and device-to-device variability. Using MemTorch~\cite{Lammie2020}, not only can we convert any \acp{DNN} to an equivalent \acp{MDNN} utilizing any memristive device model, we are also able to comprehensively investigate the effect of various device non-idealities and variation on the performance of a possible MDNN, before it is physically realized in hardware. 
	 
	In order to demonstrate an example which includes variability in our MDNN simulations, device-device variability is introduced by sampling $R_{\textnormal{OFF}}$ for each device from a normal distribution with $\bar{R}_{\textnormal{OFF}}$ = 2,500$\Omega$ with standard deviation $2\sigma$, and $R_{\textnormal{ON}}$ for each device from a normal distribution with $\bar{R}_{\textnormal{ON}}$ = 100$\Omega$ with standard deviation $\sigma$. 
	
\begin{figure}[!b]
		\centering
		\includegraphics[width=0.45\textwidth]{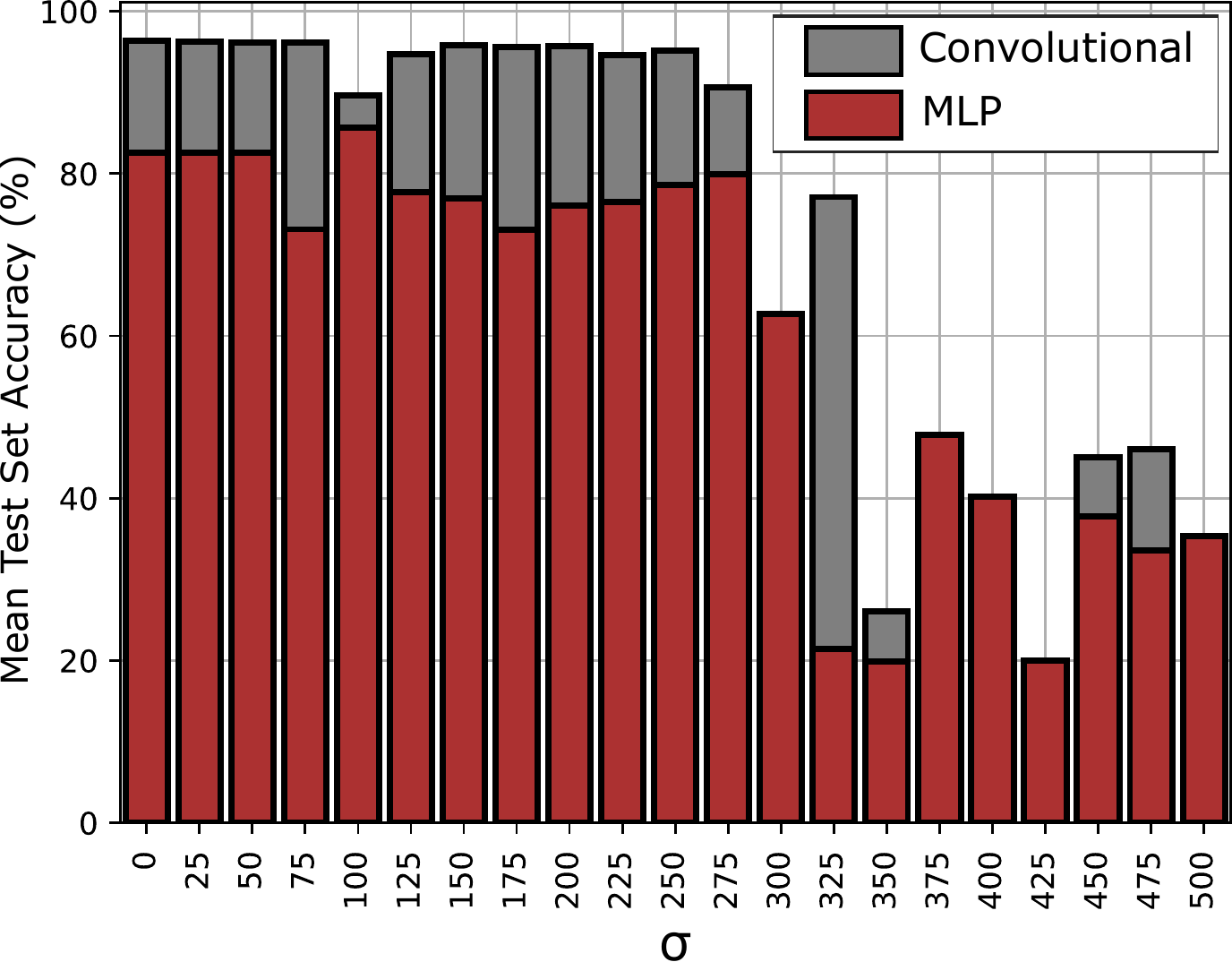}
		\caption{Simulation results investigating the performance of \acp{MDNN} for hand gesture classification adopting non-ideal Pt/Hf/Ti ReRAM devices. Device-device variability is simulated using MemTorch~\cite{Lammie2020}.}
		\label{fig:memtorch_result}
	\end{figure}

	In Fig.~\ref{fig:memtorch_result}, for the converted memristive \ac{MLP} and \ac{CNN} that process APS hand-gesture inputs, we gradually increase $\sigma$ from 0 to 500, and compare the mean test set accuracy across the three folds. As can be observed from Fig.~\ref{fig:memtorch_result}, with increasing device-to-device variability, i.e. the variability of $R_{\textnormal{ON}}$ and $R_{\textnormal{OFF}}$, the performance degradation increases across all networks. For all simulations, $R_{\textnormal{ON}}$ and $R_{\textnormal{OFF}}$ are bounded to be positive.
	
	\subsubsection{Memristive DNNs towards biomedical applications}     
	Although some previous small-scale \acp{MDNN} have been simulated for biomedical tasks such as cardiac arrhythmia classification~\cite{Hassan2018}, or have been implemented on a physical programmable memristive array for breast cancer diagnosis~\cite{Cai2019}, there is currently no large-scale \ac{MDNN}, even at the simulation-level, which has realized any practical biomedical processing tasks. 
	
	Similar to the recent advances in CMOS-driven \ac{DNN} accelerator chips discussed in subsection~\ref{subsec:cmosdnn}, there have been promises in partial~\cite{Ambrogio2018} or full~\cite{Yao2020} realizations of \acp{MDNN} in hardware, which are shown to achieve significant energy saving compared to state-of-the-art \acp{GPU}. However, unlike their CMOS counterparts, these implementations have been only able to perform simple tasks such as MNIST and CIFAR classification. This is, of course, not suitable for implementing large-scale \acp{CNN} and \acp{RNN}, which as shown in subsection~\ref{subsec:cmosdnn} are required for biomedical and healthcare tasks dealing with image~\cite{Tajbakhsh2016} or temporal~\cite{Gao2019} data types. 
	
	In addition, following similar optimization strategies as those used in CMOS accelerators, \cite{Hirtzlin2019} has simulated the use of quantized and binarized \acp{MDNN} and their error tolerance in a biomedical \ac{ECG} processing task and has shown their potential to achieve significant energy savings compared to full-precision \acp{MDNN}. However, due to the many intricacies in the design process and considering the peripheral circuitry that may offset the benefits gained by using \acp{MDNN}, full hardware design is required before the actual energy saving of such binarized \acp{MDNN} can be verified.  
	
	In the next section, we provide our analysis and perspective on the use of the three hardware technologies discussed in this section for \ac{DL}-based biomedical and healthcare applications. We also discuss how \ac{SNN}-based neuromorphic processors can benefit edge-processing for biomedical applications. 
	
    \begin{table*}[!t]
        \centering
        \caption{Existing hardware implementations and hardware-based simulations of \ac{DNN} accelerators used for healthcare and biomedical applications, and generic \ac{SNN} neuromorphic processors utilized for biomedical signal processing. $^\dagger$Simulation-based}
        \begin{tabu} to \textwidth {p{0.45\textwidth}X[l]p{0.3\textwidth}}
            \toprule
            \textbf{Biomedical or Healthcare Task} & \textbf{\ac{DNN}/\ac{SNN} Architecture} & \textbf{Hardware} \\
            \midrule
            Image-based breast cancer diagnosis \cite{mckinney2020international} & Ensemble of CNNs & CMOS (Google \ac{TPU}) \\
            Motor intention decoding~\cite{7348721} & \ac{ELM} & CMOS\\
           Spatial filtering and dimensionality reduction for brain-state classification~\cite{8490222} & Autoencoder & CMOS\\
            \midrule            
            Energy-efficient multi-class \ac{ECG} classification \cite{Corradi_etal19} & Spiking RNN & CMOS \\
            \ac{EMG} signal processing \cite{ceolini2020hand} & Spiking CNN/MLP & CMOS \\
            \ac{ECG} signal processing \cite{Bauer_etal19} & Spiking RNN & CMOS \\
            \ac{EMG} signal processing \cite{Donati_etal18} & Spiking RNN & CMOS \\
            \ac{EMG} signal processing \cite{Donati_etal19} & Feed-forward SNN & CMOS \\    \ac{EMG} and \ac{EEG} signal processing \cite{Behrenbeck_etal19} & Recurrent 3D SNN & CMOS \\ 
            \ac{EEG} and \ac{LFP} signal processing \cite{nurse2016decoding} & TrueNorth-compatible CNN & CMOS\\
            Real-time closed loop neural decoding~\cite{8717122,8854815} & Spiking ELM & CMOS\\
            \midrule
            \ac{ECG} processing for cardiac arrhythmia classification~\cite{Hassan2018} & \ac{MLP} & Memristors$^\dagger$ \\
            Breast cancer diagnosis \cite{Cai2019} & \ac{MLP} & Programmable Memristor-CMOS system \\
            \ac{ECG} signal processing~\cite{Hirtzlin2019} & Binarized CNN & Memristors$^\dagger$ \\    
            \midrule
            \ac{ECG} arrhythmia detection for hearth monitoring~\cite{Wess2017} & MLP & FPGA \\
            Mass-spectrometry for real-time cancer detection~\cite{Sanaullah2018} &  MLP & FPGA \\
            \ac{ECog} signal processing for \ac{BCI} \cite{Shrivastwa2018} & MLP & FPGA \\
            Signal processing for fall detection~\cite{5967064} & MLP & FPGA\\
            \ac{BCI}-decoding of large-scale neural sensors~\cite{Heelan2018} & LTSM & FPGA\\
            \midrule
           \ac{EEG} processing for energy-efficient Neurofeedback devices~\cite{Chen2018} & LTSM & FPGA and CMOS\\
           \ac{PPG} signal processing for heart rate estimation~\cite{9115236} & CNN/LTSM & FPGA and CMOS\\ 
           Multimodal signal classification for physical activity monitoring~\cite{8419763} & CNN & FPGA and CMOS\\
            \bottomrule
        \end{tabu} \label{tab:hwdnn}
    \end{table*}  	
	
	\section{Analysis and Perspective}\label{sec:perspective} 
    The use of \acp{ANN} trained with the backpropagation learning algorithm in the domain of healthcare and for biomedical applications such as cancer diagnosis~\cite{OhnoMachado1998} or \ac{ECG} monitoring~\cite{Ku1992} dates back to the early 90s. These networks, were typically small-scale networks run on normal workstations. As they were not deep and did not have too many parameters, they did not demand high-performance accelerators. However, with the resurgence of \acp{CNN} in the early 2010s followed by the rapid spread of \acp{DNN} and large data-sets, came the need for high-speed specialized processors. This need resulted in repurposing \acp{GPU} and actively researching other hardware and design technologies including \ac{ASIC} CMOS chips (see Table~\ref{tab:cmosdnns}) and platforms~\cite{Jouppi2017}, memristive crossbars and in-memory computing~\cite{Ambrogio2018,Yao2020,Lammie2019}, and FPGA-based designs for \ac{DNN} training~\cite{Lammie2020a,Lammie2019b} and inference~\cite{Lammie2019a}. Despite notable progress in deploying non-GPU platforms for \ac{DL} acceleration, similar to other data processing tasks, biomedical and healthcare tasks have mainly relied on standard technologies and \acp{GPU}. Currently, depending on the size of the required DNN, its number of parameters, as well as the available training dataset size, biomedical \ac{DL} tasks are usually ``trained'' on high-performance workstations with one or more GPUs~\cite{Kalaiselvi2017,Smistad2015}, on customized proprietary processors such as Google TPU~\cite{mckinney2020international}, or on various Infrastructure-as-a-Service (IaaS) provider platforms, including Nvidia GPU cloud, Google Cloud, and Amazon Web Services, among others. This is mostly due to (i) the convenience these platforms provide using high-level languages such as Python; (ii) the availability of wide-spread and open-source \ac{DL} libraries such as TensorFlow and PyTorch; and (iii) strong community and/or provider support in utilizing \acp{GPU} and IaaS for training various \ac{DNN} algorithms and applications.
    
    However, \ac{DL} inference can benefit from further research and development on emerging and mature hardware and design technologies, such as those discussed in this paper, to open up new opportunities for deploying healthcare devices closer to the edge, paving the way for low-power and low-cost \ac{DL} accelerators for \ac{PoC} devices and healthcare \ac{IoT}. Despite this fact, hardware implementations of biomedical and healthcare inference engines are very sparse. Table~\ref{tab:hwdnn} lists a summary of the available hardware implementations and hardware-based simulations of \acp{DNN} used for healthcare and biomedical signal processing applications, using the three hardware technologies covered herein. In addition, the table shows existing biomedical signal processing tasks implemented on generic low-power spiking neuromorphic processors. 
    
    \subsection{CMOS Technology Has Been the Main Player for DL Inference in the Biomedical Domain}
    Similarly to general-purpose \acp{GPU}, all other non-GPU \ac{DL} inference engines at present are implemented in CMOS. Therefore, it is obvious that most of the future edge-based biomedical platforms would rely on these inference platforms. In Table~\ref{tab:cmosdnns}, we listed a number of these accelerators that are mainly developed for low-power mobile applications. However, before the deployment of any edge-based \ac{DL} accelerators for biomedical and healthcare tasks, some challenges need to be overcome. A non-exhaustive list of these obstacles include: (i) the power and resource constraints of available mobile platforms which, despite significant improvements, are still not suitable for high-risk medical tasks; (ii) the need to verify that a \ac{DL} system can generalize beyond the distribution they are trained and tested on; (iii) bias that is inherent to datasets which may have adverse impacts on classification across different populations; (iv) confusion surrounding the liability of AI algorithms in high-risk environments \cite{eshraghian2020human}; and (v) the lack of a streamlined workflow between medical practitioners and \ac{DL}. While the latter challenges are matters of legality and policy, the former issues highlight the fundamental need to understand where dataset bias comes from, and how to improve our understanding of why neural networks learn the features they do, such that they may generalize across populations in a manner that is safe for receivers of medical care.
    
    In addition, to make the use of any accelerators possible for general as well as more complex biomedical applications, the field requires strong hardware-software co-design to build hardware that can be readily programmed for biomedical tasks. One successful co-design is the Google \ac{TPU}~\cite{Jouppi2017}, which has successfully been used to surpass human experts in medical imaging tasks~\cite{mckinney2020international}. Google has used a similar CMOS \ac{TPU} technology to design inference engines~\cite{Google2019}, which are very promising as edge hardware to enable mobile healthcare care applications. The main reason for this promise is the availability of the established software platforms (such as TensorFlow Light) and the community support for the Google \ac{TPU}.
    
    Overall, great advancements have happened for \ac{DL} accelerators in the past several years and they are currently stemming in various aspects of our life from self-driving cars to smart personal assistants. After overcoming a number of obstacles such as those mentioned above, we may be also able to widely integrate these \ac{DL} accelerators in healthcare and biomedical applications. However, for some medical applications such as monitoring that requires always-on processing, we still need systems with orders of magnitude better power efficiency, so they can run on a simple button battery for a long time. To achieve such systems, one possible approach is to process data only when available and make our processing asynchronous. A promising method to achieve such goals is the use of brain-inspired \ac{SNN}-based neuromorphic processors.  
    
    \subsection{Towards Edge Processing for Biomedical Applications With Neuromorphic Processors}
    \label{ssec:snn}
    
    Although most of the efforts presented in this work focused on \ac{DNN} accelerators, there are also notable efforts in the domain of \ac{SNN} processors that offer complementary advantages, such as the potential to reduce the power consumption by multiple orders of magnitude, and to process the data in real time. These so-called neuromorphic processors are ideal for end-to-end processing scenarios, e.g., in wearable devices where the streaming input needs to be monitored in continuous time in an always-on manner. 
    
    \begin{table*}[t]
    	\centering
        \caption{Neuromorphic platforms used for biomedical signal processing}
    	\label{tab:neurochips}
        \begin{tabu} to \textwidth {XXp{0.175\textwidth}XXX} 
            \toprule
            \textbf{Neuromorphic Chip}	& \textbf{DYNAP-SE} & \textbf{SpiNNaker} & \textbf{TrueNorth} & \textbf{Loihi} & \textbf{ODIN} \\ 
            \midrule
            \textbf{CMOS Technology}	& 180 nm & ARM968, 130 nm & 28 nm & 14 nm FinFET  & 28 nm FDSOI \\ 
            \midrule
            \textbf{Implementation}	& Mixed-signal & Digital & Digital ASIC & Digital ASIC & Digital ASIC \\ 
            \midrule
            \textbf{Neurons per core}	& 256 &  1000 (1M cores) & 256 & Max 1k  & 256 \\ 	
            \midrule
        	\textbf{Synapses per core}	& 16k & 1M & 64k & 114k-1M  & 64k  \\ 
        	\midrule
        	\textbf{Energy per SOP}	& 17 pJ @ 1.8V & Peak power 1W per chip & 26 pJ @ 0.775 & 23.6 pJ @ 0.75V & 12.7 pJ@0.55V \\
        	\midrule
        	\textbf{Size}	& 38.5 $mm^2$ & 102 $mm^2$ & - & 60 $mm^2$  & 0.086 $mm^2$  \\ 
        	\midrule
        	\textbf{Biosignal processing application}	& \ac{EMG}~\cite{Donati_etal19}, \ac{ECG}~\cite{Bauer_etal19}, \ac{HFO}~\cite{Sharifshazileh_etal19} & \ac{EMG} and \ac{EEG}~\cite{Behrenbeck_etal19} & \ac{EEG} and \ac{LFP}~\cite{nurse2016decoding} & \ac{EMG}~\cite{ceolini2020hand} &   \ac{EMG}~\cite{ceolini2020hand}\\ 
        	\bottomrule	
        \end{tabu}
    \end{table*} 
    
    There are already some works using both mixed analog-digital and digital neuromorphic platforms for biomedical tasks, showing promising results for always-on embedded biomedical systems. Table~\ref{tab:neurochips} shows a summary of today's large scale neuromorphic processors, used for biomedical signal processing. The first chip presented in this table is DYNAP-SE~\cite{Moradi_etal18}, a multi-core mixed-signal neuromorphic implementation with analog neural dynamics circuits and event-based asynchronous routing and communication. The DYNAP-SE chip has been used to implement four of the seven \ac{SNN} processing systems listed in Table~\ref{tab:hwdnn}. These \acp{SNN} are used for \ac{EMG}~\cite{Donati_etal18, Donati_etal19} and \ac{ECG}~\cite{Bauer_etal19, Corradi_etal19} signal processing. The DYNAP-SE was also used to build a spiking perceptron as part of a design to classify and detect High-Frequency Oscillations (HFO) in human intracranial \ac{EEG}~\cite{Sharifshazileh_etal19}.

   In \cite{Corradi_etal19,Bauer_etal19,Donati_etal18} a spiking \ac{RNN} is used to integrate the \ac{ECG}/\ac{EMG} patterns temporally and separate them in a linear fashion to be classifiable with a linear read-out. A \ac{SVM} and linear least square approximation is used in the read out layer for \cite{Bauer_etal19,Corradi_etal19} and overall accuracy of $91\%$ and $95\%$ for anomaly detection were reached respectively. In \cite{Donati_etal18}, the timing and dynamic features of the spiking \ac{RNN} on \ac{EMG} recordings was investigated for classifying different hand gestures. In~\cite{Donati_etal19} the performance of a feedforward \ac{SNN} and a hardware-friendly spiking learning algorithm for hand gesture recognition using superficial \ac{EMG} was investigated and compared to traditional machine learning approaches, such as \ac{SVM}. Results show that applying \ac{SVM} on the spiking output of the hidden layer achieved a classification rate of $84\%$, and the spiking learning method achieved $74\%$ with a power consumption of about $0.05~mW$. This was compared to state-of-the-art embedded system showing that the proposed spiking network is two orders of magnitude more power efficient~\cite{benatti2015versatile, montagna2018pulp}.
    
    The other neuromorphic platforms listed in Table~\ref{tab:neurochips} include digital architectures such as SpiNNaker~\cite{Furber2013}, TrueNorth~\cite{Merolla2004} and Loihi~\cite{Davies_etal18}. SpiNNaker has been used for \ac{EMG} and \ac{EEG} processing and the results show improved classification accuracy compared to traditional machine learning methods~\cite{Behrenbeck_etal19}. In~\cite{nurse2016decoding}, the authors developed a framework for decoding \ac{EEG} and \ac{LFP} using \acp{CNN}. The network was first developed in Caffe and the result was then used as a basis for building a TrueNorth-compatible neural network. The TrueNorth-compatible network achieved the highest classification, at approximately 76\%. In~\cite{8717122,8854815}, the authors present a low-power neuromorphic platform named Spike-input Extreme Learning Machine (SELMA), which performs  continuous state decoding towards fully-implantable wireless intracortical \ac{BMI}. Recently, the benchmark hand-gesture classification introduced in subsection \ref{subsec:benchmark}, was processed and compared on two additional digital neuromorphic platforms, Loihi and ODIN/MorphIC~\cite{Frenkel_etal18, frenkel_etal19}. A spiking \ac{CNN} was implemented on Loihi and a spiking \ac{MLP} was implemented on ODIN/MorphIC~\cite{ceolini2020hand}. The results achieved using these networks are presented in Table~\ref{tab:comparison}.    

         On-chip adaptation and learning mechanisms, such as those present in some of the neuromorphic devics  listed in Table~\ref{tab:neurochips}, could be a game changer for personalized medicine, where the system can adapt to each patient's unique bio signature and/or drift over time.
    However, the challenge of implementing efficient on-chip online learning in these types of neuromorphic architectures has not yet been solved. This challenge lies on two main factors: \emph{locality} of the weight update and \emph{weight storage}. 
    \paragraph*{Locality} There is a hardware constraint that the learning information for updating the weights of any on-chip network should be locally available to the synapse, otherwise most of the silicon area would be consumed by the wires, required to route the update information to it. As Hebbian learning satisfies this requirement, most of the available on-chip learning algorithms focus on its implementation in forms of unsupervised/semi-supervised learning \cite{Frenkel_etal18,qiao_etal5_rolls}.  
    However, local Hebbian-based algorithms are limited in learning static patterns or using very shallow networks~\cite{RahimiAzghadi2015}. 
    There are also some efforts in the direction of on-chip gradient-descent based methods which implement on-chip error-based learning algorithms where the least mean square of a neural network cost function is minimized. 
    For example, spike-based delta rule is the most common weight update used for single-layer networks which is the base of the back-propagation algorithm used in the vast majority of current multi-layer neural networks. Single layer mixed-signal neuromorphic circuit implementation of the delta rule have already been designed \cite{payvand_indiveri_19} and employed for \ac{EMG} classification \cite{Donati_etal19}. Expanding this to multi-layer networks involves non-local weight updates which limits its on-chip implementation. Making the backpropagation algorithm local is a topic of on-going research \cite{kaiser_etal20_decolle,bellec2019_eprop,sacramento_etal18_dendritic}. 
    
    \paragraph*{Weight storage} The holy grail weight storage for online on-chip learning is a memory with non-volatile properties whose state can change linearly in an analog fashion. Non-volatile memristive devices provide a great potential for this. Therefore, there is a large body of literature in combining the maturity of CMOS technology with the potential of the emerging memories to take the best out of the two worlds. 
    
    The integration of CMOS technology with that of the emerging devices has been demonstrated for non-volatile filamentary switches \cite{valentian2019IEDM} already at a commercial level \cite{hayakawa2015VLSI}. There have also been some efforts in combining CMOS and memristor technologies to design supervised local error-based learning circuits using only one network layer by exploiting the properties of memristive devices \cite{payvand_indiveri_19,dalgaty_etal_2019_AIP,payvand_etal20_icc}.	
    
    Apart from the above-mentioned benefits in utilizing memristive devices for online learning in \ac{SNN}-based neuromorphic chips, as discussed in subsection~\ref{subsec:memdnn}, memristive devices have also shown interesting features to improve the power consumption and delay of conventional \acp{DNN}. However, as shown in Table~\ref{tab:hwdnn}, memristor-based \acp{DNN} are very sparse in the biomedical domain, and existing works are largely based only on simulation.  
    
    \subsection{Why Is the Use of MDNNs Very Limited in the Biomedical Domain?}
    Currently there are very few hardware implementations of biomedical \acp{MDNN} that make use of general programmable memristive-CMOS, and only one programmed to construct an \ac{MLP} for cancer diagnosis. We could also find two other memristive designs in literature for biomedical applications (shown in Table~\ref{tab:hwdnn}), but they are only simulations considering memristive crossbars. This sparsity is despite the significant advantages that memristors provide in \ac{MAC} parallelization and in-memory computing paradigm, while being compatible with CMOS technology~\cite{chicca2020recipe}. These features make memristors ideal candidates for \ac{DL} accelerators in general, and for portable and edge-based healthcare applications in particular, because they have stringent device size and power consumption requirements. To be able to use memristive devices in biomedical domain, though, several of their shortcomings such as limited endurance, mismatch, and analog noise accumulation must be overcome first. This demands further research in the materials, as well as the circuit and system design side of this emerging technology, while at the same time developing facilitator open-source software~\cite{Lammie2020} to support \acp{MDNN}. Furthermore, investigating the same techniques utilized in developing CMOS-based \ac{DL} accelerators such as limited precision data representation~\cite{Lammie2019,Hirtzlin2019} and approximate computing schemes can lead to advances in developing \acp{MDNN} and facilitate their use in biomedical domains. 
    
	\subsection{Why and When to Use FPGA for Biomedical DNNs?} 
	Table~\ref{tab:hwdnn} shows that \ac{FPGA} is a popular hardware technology for implementing simple \ac{DL} networks such as \acp{MLP} \cite{Wess2017,Sanaullah2018,Shrivastwa2018,5967064} and in a few cases, more complex \acp{LSTM} and \acp{CNN} \cite{Chen2018,Heelan2018,9115236,8419763}. The table also shows that \acp{FPGA} are mainly used for signal processing tasks and have not been widely used to run complex \ac{DL} architectures such as \acp{CNN}. This is mainly because they have limited on-chip memory and low bandwidth compared to GPUs. However, they demonstrate notable benefits in terms of significantly shorter development time compared to \acp{ASIC}, and much lower power consumption than typical \acp{GPU}. Besides, significant power and latency improvement can be gained by customizing the implementation of various components of a \ac{DNN} on an \ac{FPGA}, compared to running it on a general-purpose CPU or GPU~\cite{Sanaullah2018,Chen2018}. For instance, in~\cite{Chen2018}, \ac{EEG} signals are processed on \acp{FPGA} using two customized hardware blocks for (i) parallelizing \ac{MAC} operations and (ii) efficient recurrent state updates, both of which are key elements of \acp{LSTM}. This has resulted in almost an order of magnitude power efficiency compared to \acp{GPU}. This efficiency is critical in many edge-computing applications including DNN-based point-of-care biomedical devices~\cite{Xie2019} and healthcare \ac{IoT}~\cite{Farahani2020,Azimi2018}. 
	
	Another benefit of FPGAs is that a customized efficient FPGA design can be directly synthesized into an \ac{ASIC} using a nanometer-node CMOS technology to achieve even more benefits~\cite{9115236,8419763}. For instance, \cite{Chen2018} has shown near 100 times energy efficiency improvement as an \ac{ASIC} in a 15-nm CMOS technology, compared to its \ac{FPGA} counterpart.
	
	Although low-power consumption and affordable cost are two key factors for almost any edge-computing or near-sensor device, these are even more important for biomedical devices such as wearables, health-monitoring systems, and PoC devices. Therefore, \acp{FPGA} present an appealing solution, where their limitations can be addressed for a customized \ac{DNN} using specific design methods such as approximate computing~\cite{Lammie2019b} and limited-precision data~\cite{Lammie2019a,Lammie2020a}, depending on the cost, required power consumption, and the acceptable accuracy of the biomedical device.  

	Another programmable low-power device that can be used in biomedical applications are \acp{FPAA}. These are constructed using programmable \acp{CAB} and interconnects. Unlike \acp{FPGA}, \acp{FPAA} tend to be more application driven than general purpose as they may be current mode or voltage mode devices~\cite{1329340}. 
	FPAAs have been shown to perform computation with 1000 times more power efficiency while reducing the required area by 100 times when compared to FPGAs~\cite{7591797}. Therefore, they are a promising candidate for accelerating biomedical signal processing if machine learning algorithms such as \acp{ANN} can be implemented using them.   
	
	In 2003, \cite{1275780} explored \acp{ANN} with differential feedback, and in 2006 \cite{1716286} implemented an \ac{ANN} using multi-chip \acp{FPAA}. More recently, \cite{6032715} have demonstrated that \acp{VMM} can be efficiently computed using \acp{FPAA}, which can be used to compute linear and unrolled convolution layers within \acp{DNN}. However, while \acp{FPAA} have been used in several biomedical applications ranging from knee-joint rehabilitation~\cite{7591797} to the amplification of various bio-electric signals~\cite{5709584}, the implementation of a \ac{FPAA} \ac{DNN} accelerator, which can be used in biomedical and general applications, is yet to be explored.

    \begin{table*}[!t]
        \centering
        \caption{Comparison of conventional \acp{DNN} implemented on various hardware platforms with spiking \ac{DNN} neuromorphic systems on the benchmark biomedical signal processing task of hand gesture recognition for both single sensor and sensor fusion, as explained in subsection~\ref{subsec:benchmark}. The results of the accuracy are reported with mean and standard deviation obtained over a 3-fold cross validation. Loihi, Embedded GPU, and ODIN+MorphIC implementation results are from~\cite{ceolini2020hand}. The \ac{DNN} architectures adopted are as follows: $^\diamond$8c3-2p-16c3-2p-32c3-512-5 CNN. $^\dagger$16-128-128-5 MLP. $^\ddagger$16-230-5 MLP. $^\mp$4 $\times$ 400-210-5 MLP. $^\cup$EMG and APS/DVS networks are fused using a 5-neuron dense layer.}
        \begin{tabu} to \textwidth {>{\centering}p{0.15\textwidth}p{0.15\textwidth}XX[r]X[r]X[r]}
            \toprule
            \multicolumn{1}{c}{ \textbf{Platform}} & \textbf{Modality}  & \textbf{Accuracy (\%)} & \textbf{Energy (uJ)}  & \textbf{Inference time (ms)}  & \textbf{EDP (uJ * s)}  \\ 
            \midrule
            \multirow{3}{*}{\begin{tabular}[c]{@{}c@{}} \textbf{Loihi}\\(Spiking) \end{tabular}} & EMG (MLP$^\dagger$)  & 55.7 $\pm$ 2.7  & 173.2 $\pm$ 21.2  & 5.89 $\pm$ 0.18  & 1.0 $\pm$ 0.1  \\
 & DVS (CNN$^\diamond$)  & 92.1 $\pm$ 1.2  & 815.3 $\pm$ 115.9  & 6.64 $\pm$ 0.14  & 5.4 $\pm$ 0.8  \\
 & EMG+DVS (CNN$^\cup$)  & 96.0 $\pm$ 0.4  & 1104.5 $\pm$ 58.8  & 7.75 $\pm$ 0.07  & 8.6 $\pm$ 0.5  \\ 
\midrule
\multirow{3}{*}{\begin{tabular}[c]{@{}c@{}}\textbf{ODIN+MorphIC}\\(Spiking) \end{tabular}} & EMG (MLP$^\ddagger$) & 53.6~$\pm$~1.4~~ & 7.42~$\pm$~0.11 & 23.5~$\pm$~0.35 & 0.17~$\pm$~0.01 \\
 & DVS (MLP$^\mp$) & 85.1~$\pm$~4.1 & 57.2~$\pm$~6.8 & 17.3~$\pm$~2.0 & 1.00~$\pm$~0.24 \\
 & EMG+DVS (MLP$^\cup$)~ & 89.4~$\pm$~3.0 & 37.4~$\pm$~4.2 & 19.5~$\pm$~0.3 & 0.42~$\pm$~0.08 \\ 
\midrule
\multirow{6}{*}{\textbf{Embedded GPU}~ } & EMG (MLP$^\dagger$) & 68.1~$\pm$~2.8 & (25.5~$\pm$~8.4)~$\cdot10^3$ & 3.8~$\pm$~0.1 & 97.3~$\pm$~4.4 \\
\multicolumn{1}{l}{} & EMG (MLP$^\ddagger$) & 67.2~$\pm$~3.6 & (23.9~$\pm$~5.6)~$\cdot 10^3$ & 2.8~$\pm$~0.08 & 67.2~$\pm$~2.9 \\
\multicolumn{1}{l}{} & APS (CNN$^\diamond$) & 92.4~$\pm$~1.6 & (31.7~$\pm$~7.4)~$\cdot 10^3$ & 5.9~$\pm$~0.1 & 186.9~$\pm$~3.9 \\
\multicolumn{1}{l}{} & APS (MLP$^\mp$) & 84.2~$\pm$~4.3 & (30.2~$\pm$~7.5)~$\cdot 10^3$ & 6.9~$\pm$~0.1 & 211.3~$\pm$~6.1 \\
\multicolumn{1}{l}{} & EMG+APS (CNN$^\cup$) & 95.4~$\pm$~1.7 & (32.1~$\pm$~7.9)~$\cdot 10^3$ & 6.9~$\pm$~0.05 & 221.1~$\pm$~4.1 \\
\multicolumn{1}{l}{} & EMG+APS (MLP$^\cup$) & 88.1~$\pm$~4.1 & (32.0~$\pm$~8.9)~$\cdot 10^3$ & 7.9~$\pm$~0.05 & 253~$\pm$~3.9 \\ 
\midrule
\multirow{5}{*}{\textbf{FPGA} } & EMG (MLP$^\dagger$) & 67.2~$\pm$~2.3 & (17.6~$\pm$~1.1)~$10^3$ & 4.2~$\pm$~0.1 & 74.1~$\pm$~1.2 \\
 & EMG (MLP$^\ddagger$) & 63.8~$\pm$~1.4 & (13.9~$\pm$~1.8)~$\cdot 10^3$ & 3.5~$\pm$~0.1 & 48.9~$\pm$~1.9 \\
 & APS (CNN$^\diamond$) & 96.7~$\pm$~3.0 & (24.0~$\pm$~1.2)~$10^3$ & 5.4~$\pm$~0.2 & 130.8~$\pm$~1.4 \\
 & APS (MLP$^\mp$) & 82.9~$\pm$~8.4 & (23.1~$\pm$~2.6)~$\cdot 10^{3}$~~ & 5.7~$\pm$~0.2 & 131.4~$\pm$~2.8 \\
 & EMG+APS (CNN$^\cup$) & 94.8~$\pm$~2.0 & (31.2~$\pm$~3.0)~$10^3$ & 6.3~$\pm$~0.1 & 196.3~$\pm$~3.1 \\ 
 & EMG+APS~(MLP$^\cup$) & 83.4 $\pm$ 2.8  & (31.1 $\pm$ 1.4) $\cdot 10^3$  & 7.3 $\pm$ 0.2  & 228.2 $\pm$ 1.6  \\ 
\midrule
\multirow{6}{*}{\textbf{Memristive} } & EMG (MLP$^\dagger$) & 64.6~$\pm$~2.2 & 0.038 & 6.0~$\cdot10^{-4}$~~ & 2.38~$\cdot10^{-8}$~~ \\
 & EMG (MLP$^\ddagger$) & 63.8~$\pm$~1.4 & 0.026 & 4.0~$\cdot10^{-4}$ & 1.04~$\cdot10^{-8}$ \\
 & APS (CNN$^\diamond$) & 96.2~$\pm$~3.3 & 4.83 & 1.0~$\cdot10^{-3}$ & 4.83~$\cdot10^{-6}$ \\
 & APS (MLP$^\mp$) & 82.4~$\pm$~8.5 & 0.18 & 4.0~$\cdot10^{-4}$ & 7.2~$\cdot10^{-8}$ \\
 & EMG+APS (CNN$^\cup$) & 94.8~$\pm$~2.0 & 4.90 & 1.2~$\cdot10^{-3}$ & 5.88~$\cdot10^{-6}$ \\
 & EMG+APS (MLP$^\cup$) & 83.4~$\pm$~2.8 & 0.33 & 6.0~$\cdot10^{-4}$ & 1.98~$\cdot10^{-7}$ \\
            \bottomrule
        \end{tabu} \label{tab:comparison}
    \end{table*}
	
	\subsection{Benchmarking EMG Processing Across Multiple DNN and SNN Hardware Platforms}
	In Table~\ref{tab:comparison}, we compare our FPGA and memristive implementations to other \ac{DNN} accelerators and neuromorphic processors from~\cite{ceolini2020hand}. In~\cite{ceolini2020hand}, the authors presented a sensor fusion neuromorphic benchmark for hand-gesture recognition based on EMG and event-based camera. Two neuromorphic platforms, Loihi~\cite{Davies_etal18} and ODIN+MorphIC~\cite{Frenkel_etal18,frenkel_etal19}, were deployed and the results were compared to traditional machine learning baselines implemented on an embedded GPU, the NVIDIA Jetson Nano. Loihi and ODIN+MorphIC are digital neuromorphic platforms. Loihi is a 128-core neuromorphic chip fabricated on 14\,nm FinFET process, designed by Intel Labs. It implements adaptive self-modifying event-driven fine-grained parallel computations used to implement learning and inference with high efficiency. ODIN (Online-learning Digital spiking Neuromorphic) is designed using 28\,nm FDSOI CMOS technology and consists of a single neurosynaptic core with 256 neurons and 256$^2$ synapses that embed a 3-bit weight and a mapping table bit that allows enabling or disabling \ac{STDP}. MorphIC is a quad-core digital neuromorphic processor with 2k \ac{LIF} neurons and more than 2M synapses in 65\,nm CMOS technology \cite{frenkel_etal19}. They can be either programmed with offline-trained weights or trained online with a stochastic version of \ac{SDSP}. 
	
	For the spiking architectures shown in Table~\ref{tab:comparison}, the vision input and \ac{EMG} data were individually processed using spiking \ac{CNN} and spiking \ac{MLP} respectively, and fused in the last layer. Loihi was trained using SLAYER~\cite{shrestha2018slayer}, a backpropagation framework used for evaluating the gradient of any kind of \ac{SNN}. It is a dt-based \ac{SNN} backpropagation algorithm that keeps track of the internal membrane potential of the spiking neuron and uses it during gradient propagation. Both ODIN and Morphic training was carried out in Keras with quantization-aware stochastic gradient descent following a standard ANN-to-SNN mapping approach.
	
	The dataset used is described in Section~\ref{subsec:benchmark}. It is a collection of 5 hand gestures from sign language (e.g. ILY)\footnote{\url{https://zenodo.org/record/3663616\#.X2m5GC2cbx4}. Further implementation details can be found in \cite{ceolini2020hand}.}. In the comparison proposed in Table~\ref{tab:comparison} the input and hidden layers are sequenced with the ReLU activation function, and output layers are fed through Softmax activation functions to determine class probabilities. Dropout layers are used in all networks to avoid over-fitting. The \ac{DNN} architectures are determined in the table caption.   

	The platforms used for each system in Table~\ref{tab:comparison} are as follows: ODIN+MorphIC~\cite{Frenkel_etal18, frenkel_etal19} and Loihi~\cite{Davies_etal18} neuromorphic platforms were used for spiking implementations; NVIDIA Jetson Nano was used for all embedded GPU implementations; OpenVINO Toolkit FPGA was used for all FPGA implementations, and MemTorch~\cite{Lammie2020} was used for converting the MLP and CNN networks to their corresponding \acp{MDNN} to determine the test set accuracies of all memristive implementations.  
	
	From Table~\ref{tab:comparison}, it can be observed that, when transitioning from generalized architectures to application specific processors, more optimized processing of a subset of given tasks can be achieved. Moving up the specificity hierarchy from GPU to FPGA to memristive networks shows orders of magnitude of improvement in both MLP and CNN processing, but naturally at the expense of a generalizable range of tasks. While GPUs are relatively efficient at training networks (compared to CPUs), the impressive metrics presented by memristor (\ac{RRAM} in this simulations) is coupled with limited endurance. This is not an issue for read-only tasks, as is the case with inference, but training is thwarted by the thousands of epochs of weight updates which limits broad use of \acp{RRAM} in training. Rather, more exploration in alternative resistive-based technologies such as \ac{MRAM} could prove beneficial for tasks that demand high endurance. 
	
	After determining the test set accuracy of each \ac{MDNN} using MemTorch~\cite{Lammie2020}, we determined the energy required to perform inference on a single input, the inference time, and the \ac{EDP} by adopting the metrics in~\cite{wang2019deep}, for a tiled memristor architecture. All assumptions made in our calculations are listed below. Parameters are adopted from those given in a 1T1R 65nm technology, where the maximum current during inference is 3$\rm\mu$A per cell with a read voltage of 0.3V. Each cell is capable of storing 8 bits with a resistance ratio of 100, and mapping signed weights is achieved using a dual column representation. All convolutions are performed by unrolling the kernels and performing MVMs, and the fully connected layers have the fan-in weights for a single neuron assigned to one column. Each crossbar has an aspect ratio of 256$\times$64 to enable more analog operations per ADC when compared to a 128$\times$128 array. Where there is insufficient space to map weights to a single array, they are distributed across multiple arrays, with their results to be added digitally. Throughput can be improved at the expense of additional arrays for convolutional layers, by duplicating kernels such that multiple inputs can be processed in parallel. The number of tiles used for each network is assumed to be the exact number required to balance the processing time of each layer. The power consumption of each current-mode 8-bit ADC is estimated to be 2$\times$10$^{-4}$ W with an operating frequency of 40 MHz (5 MHz for bit-serial operation) \cite{wang2019deep}. The ADC latency is presumed to dominate digital addition of partial products from various tiles. The dynamic range of each ADC has been adapted to the maximum possible range for each column, and each ADC occupies a pair of columns.  
	
	The above presumptions lead to pre-silicon results that are extremely promising for memristor arrays, as shown in Table~\ref{tab:comparison}. But it should be clear that these calculations were performed for \textit{network-specific} architectures, rather than a more general application-specific use-case. That is, we assume the chip has been designed for a given neural network model. The other comparison benchmarks are far more generalizable, in that they are suited to not only handle most network topologies, but are also well-suited for training. The substantial improvement of inference time over other methods is a result of duplicate weights being mapped to enable higher parallelism, which is tolerable for small architectures, but lends to prohibitively large ADC power consumption for computer vision tasks which rely on deep networks and millions of parameters, such as VGG-16. In addition, the area of each ADC is estimated to be 3$\times$10$^{-3}$mm$^2$, which is orders of magnitude larger than the area of each RRAM cell (1.69$\times$10$^{-7}$mm$^2$). This disparity implies that pitch-matching is not viable. Instead, to achieve parallelism, weights must be duplicated across tiles which demands redundancy. This improvement in parallelism thus comes at the cost of additional area and power consumption.
	The use of memristors as synapses in spike-based implementations may be more appropriate, so as to reduce the ADC overhead by replacing multi-bit ADCs with current sense amplifiers instead, and reducing the reliance on analog current summation along resistive and capacitive bit-lines.   
    
	Spike-based hardware show approximately two orders of magnitude improvement in the EDP from Table~\ref{tab:comparison} when compared to their GPU and FPGA counterparts, which highlights the prospective use of such architectures in always-on monitoring. This is necessary for enhancing the prospect of ambient-assisted living, which would allow medical resources to be freed up for tasks that are not suited for automation. In general, one would expect that data should be processed in its naturalized form. For example, 2D \acp{CNN} do not discard the spatial relations between pixels in an image. Graph networks are optimized for connectionist data, such as the structure of proteins. By extension, the discrete events generated by electrical impulses such as in \acp{EMG}, \acp{EEG} and \acp{ECG} may also be optimized for \acp{SNN}. Of course, this discounts any subthreshold firing patterns of measured neuron populations. But one possible explanation for the suitability of spiking hardware for biological processes stems from the natural timing of neuronal action potentials. Individual neurons will typically not fire in excess of 100 Hz, and the average heart rate (and correspondingly, ECG spiking rate) will not exceed 3 Hz. There is a clear mismatch between the clock rate of non-spiking neural network hardware, which tend to at least be in the MHz range, and spike-driven processes. This introduces a significant amount of wastage in processing data when there is no new information to process (e.g., in between heartbeats, action potentials, or neural activity). 
	
	Nonetheless, it is clear that accuracy is compromised when relying on \ac{EMG} signals alone, based on the approximately 10\% decrease of classification accuracy on the Loihi chip and ODIN+MorphIC, as against their GPU/FPGA counterparts. 
	This could be a result of spike-based training algorithms lagging behind in maturity compared to conventional neural network methods, or it could be an indication that critical information is being discarded when neglecting the subthreshold signals generated by populations of neurons. But when \ac{EMG} and \ac{DVS} data are combined, this multi-sensory data fusion of spiking signals positively reinforce upon each other with an approximately 4\% accuracy improvement, whereas combining non-spiking, mismatched data representations leads to marginal improvements, and even a destructive effect (e.g., non-spiking CNN implementation on FPGA and memristive arrays). This may be a result of EMG and APS data taking on completely different structures. This is a possible indication that feature extraction from merging the same structural form of data (i.e., as spikes) proves to be more beneficial than combining a pair of networks with two completely different modes of data (i.e., EMG signals with pixel-driven images). This allows us to draw an important hypothesis: neural networks can benefit from a consistent representation of data generated by various sensory mechanisms. This is supported by biology, where all biological interpretations are typically represented by graded or spiking action potentials.
	
	\subsection{Deep Network Accelerators and Patient-specific Model Tuning} \label{sec:patientspecific}
	Given the inherent variability between patients, it is difficult to train and deploy a single model to a large group of individuals each with unique signature(s). Consequently, significant efforts are being made to facilitate patient-specific model tuning processes ~\cite{1403183,7302516,9040275}. \ac{PSM} is the development of computational models of human or animal pathophysiology that are individualized to patient-specific data~\cite{1403183}. 
	
	In the \ac{DL} domain, existing \ac{ANN} and neuromorphic models can be retrofitted to specific patients using transfer learning and tuning algorithms. In this approach, the network is first trained on a large dataset including data from various patients to acquire the domain-specific knowledge of the targeted task. Parts of the large network are then retrained, i.e. tuned, using patient-specific data, to produce better performance for individual patients. This way, the domain-specific features of the large network are transferred to the smaller network that is retrained to learn patient-specific features \cite{9040275}. Depending on the availability of patient-specific data, \ac{PSM} can be performed online (on-chip) or offline (off-chip). 
	
	\subsubsection{Online patient-specific model tuning}
	Considering concerns surrounding the sensitive nature of individual patient data, and the ability of some recent edge-AI CMOS chips such as LNPU~\cite{Lee2019} to perform online training, patient-specific model tuning can be performed online on the hardware deep learning accelerator. To achieve this, a sufficient amount of patient data that is fed to the accelerator over time can be gathered to individualize the initial generic model. An accelerator that can adapt its working to the specific needs of a patient would be highly beneficial but it may require buffering of data~\cite{yoo20128}, which needs higher on-chip memory and may introduce power overheads.      
	
	\subsubsection{Offline patient-specific model tuning}
    A convenient approach to tune general models, with domain-specific knowledge, to patient-specific data is offline off-chip transfer learning. However, unlike online tuning, the offline approach requires prior patient data measurements, which may not be readily available. Besides, the offline approach may require undesired remote storage and processing of private patient data to retrain and tune generic models. 
	
	\section{Conclusion} \label{sec:conclusion}
	The use of \ac{DL} in biomedical signal processing and healthcare promises significant utility for medical practitioners and their patients. \acp{DNN} can be used to improve the quality of life for chronically ill patients by enabling ambient monitoring for abnormalities, and correspondingly can reduce the burden on medical resources. Proper use can lead to reduced workloads for medical practitioners who may divert their attention to time-critical tasks that require a standard beyond what neural networks can achieve at this point in time. 
	
	We have stepped through the use of various \ac{DL} accelerators on a disparate range of medical tasks, and shown how \acp{SNN} may complement \acp{DNN} where hardware efficiency is the primary bottleneck for widespread integration. We have provided a balanced view to how memristors may lead to optimal hardware processing of both \acp{DNN} and \acp{SNN}, and have highlighted the challenges that must be overcome before they can be adopted at a large-scale. While the focus of this tutorial and review is on hardware implementation of various \ac{DL} algorithms, the reader should be mindful that progress in hardware is a necessary, but insufficient, condition for successful integration of medical-AI.
	
	Adopting medical-AI tools is clearly a challenge that demands the collaborative attention of healthcare providers, hardware and software engineers, data scientists, policy-makers, cognitive neuroscientists, device engineers and materials scientists, amongst other specializations. A unified approach to developing better hardware can have pervasive impacts upon the healthcare industry, and realize significant payoff by improving the accessibility and outcomes of healthcare. 
	
    \section*{Acknowledgment}
    M. Rahimi Azghadi acknowledges a JCU Rising Start ECR Fellowship. C. Lammie acknowledges the JCU DRTPS. This paper is supported in part by the European Union’s Horizon 2020 ERC project NeuroAgents (Grant No. 724295). It was also partially funded by EU H2020 grants 824164 ``HERMES", 871371 ``Memscales”, PCI2019-111826-2 ``APROVIS3D”, and by Spanish grant from the Ministry of Science and Innovation PID2019-105556GB-C31 (NANOMIND) (with support from the European Regional Development Fund).


\vfill		
\begin{IEEEbiography}[{\includegraphics[width=1in,height=1.25in,clip,keepaspectratio]{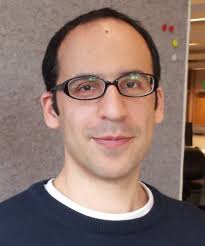}}]{Mostafa Rahimi Azghadi}
	(S'07--M'14--SM'19) completed his PhD in Electrical \& Electronic Engineering at The University of Adelaide, Australia, earning the Doctoral Research Medal, as well as the Adelaide University Alumni Medal.
	He is currently a senior lecturer in the College of Science and Engineering, James Cook University, Townsville, Australia, where he researches low-power and high-performance neuromorphic accelerators for neural-inspired and deep learning networks for a variety of applications including automation, precision agriculture, aquaculture, marine sciences, and medical imaging. His research has attracted over \$0.7M in funding from national and international resources.  
	
	Dr. Rahimi was a recipient of several national and international accolades including a 2015 South Australia Science Excellence award, a 2016 Endeavour Research Fellowship, a 2017 Queensland Young Tall Poppy Science award, a 2018 JCU Rising Star ECR Leader fellowship, and a 2019 Fresh Science Queensland finalist. Dr Rahimi is a senior member of the IEEE and a TC member of Neural Systems and Applications of the circuit and system society. He serves as an associate editor of {\it Frontiers in Neuromorphic Engineering} and {\it IEEE Access}.
\end{IEEEbiography}

\newpage

\begin{IEEEbiography}[{\includegraphics[width=1in,height=1.2in,clip,keepaspectratio]{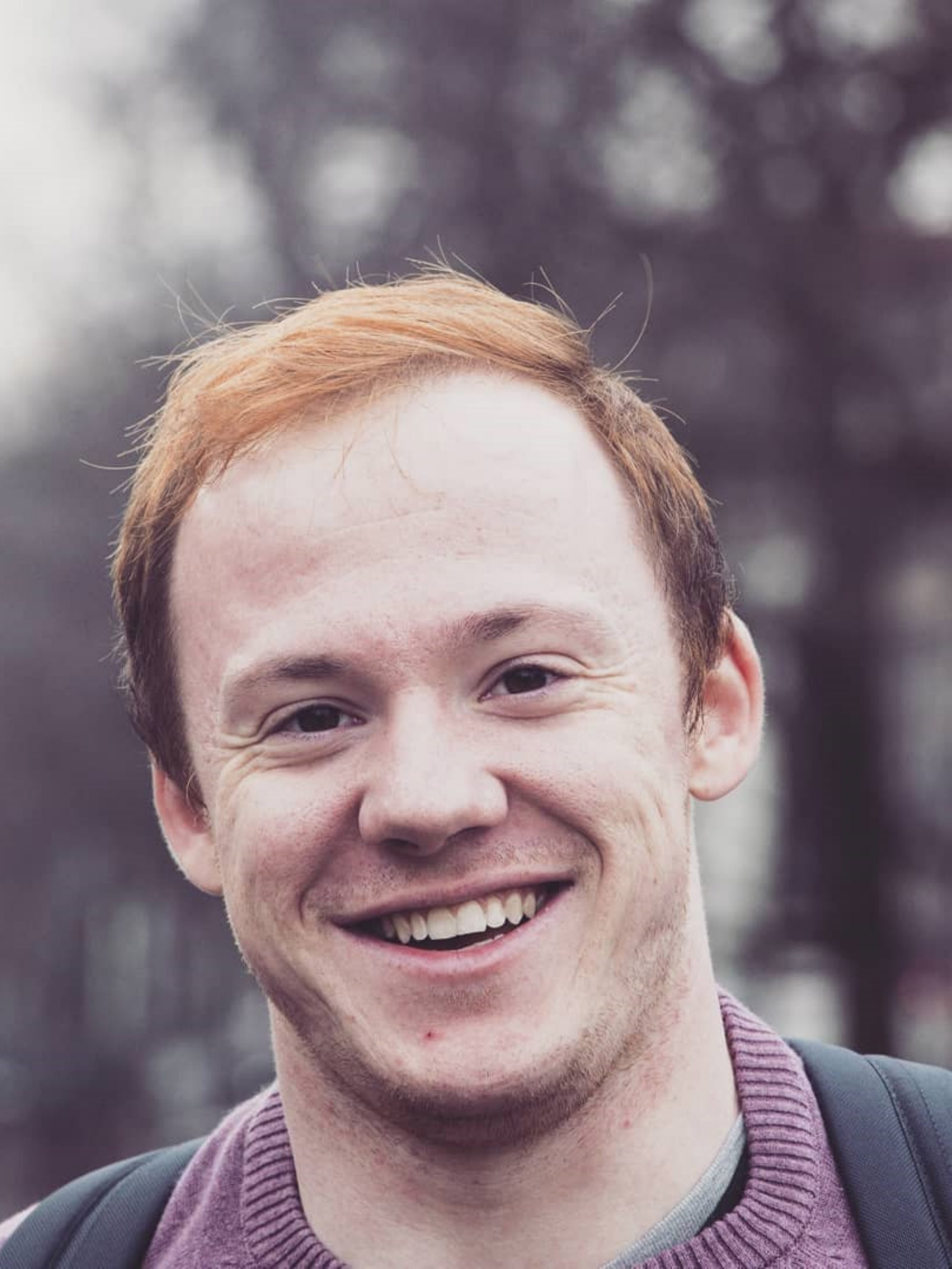}}]{Corey Lammie} 
    (S'17) is currently pursuing a PhD in Computer Engineering at James Cook University (JCU), where he completed his undergraduate degrees in Electrical Engineering (Honours) and Information Technology in 2018. His main research interests include brain-inspired computing, and the simulation and hardware implementation of Spiking Neural Networks (SNNs) and Artificial Neural Networks (ANNs) using ReRAM devices and FPGAs. He has received several awards and fellowships including the intensely competitive 2020-2021 IBM international PhD Fellowship, a Domestic Prestige Research Training Program Scholarship, and the 2017 Engineers Australia CN Barton Medal awarded to the best undergraduate engineering thesis at JCU. Corey has served as a reviewer for several IEEE journals and conferences including IEEE Transactions on Circuits and Systems and the IEEE International Symposium on Circuits and Systems (ISCAS).
\end{IEEEbiography}

\vspace{-2cm}

\begin{IEEEbiography}[{\includegraphics[width=1in,height=1.2in,clip,keepaspectratio]{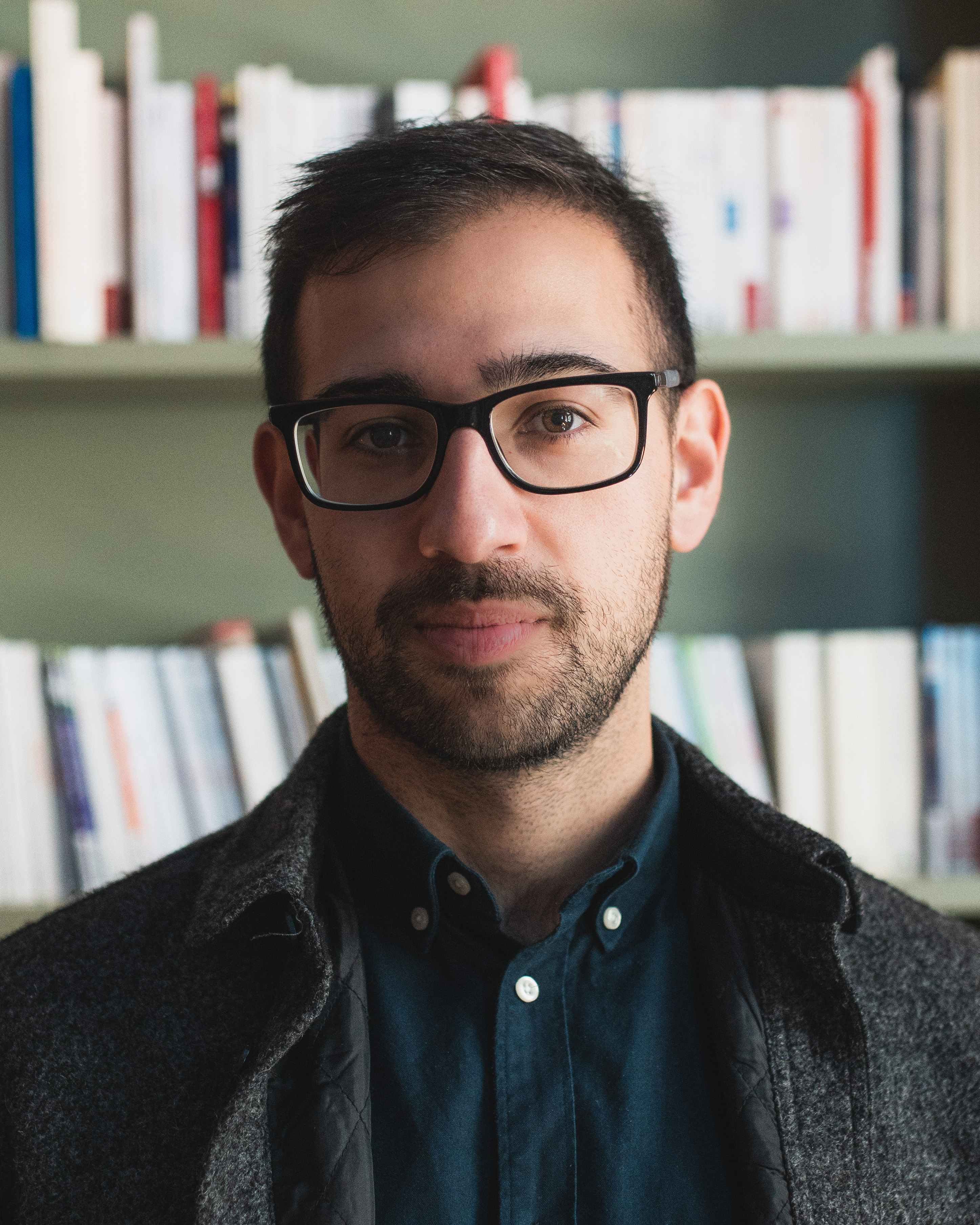}}]{Jason K. Eshraghian} 
    (M'19-S'16) is a Post-Doctoral Researcher at the Department of Electrical Engineering and Computer Science, University of Michigan in Ann Arbor. He received the Bachelor of Engineering (Electrical \& Electronic) and the Bachelor of Laws degrees from The University of Western Australia, Perth, WA, Australia in 2016, where he also completed his Ph.D. Degree. From 2015 to 2016, he was a Research Assistant at Chungbuk National University, South Korea. He is a member of the IEEE Neural Systems and Applications Committee. His current research interests include neuromorphic computing and spiking neural networks. Jason was awarded the 2019 IEEE Very Large Scale Integration Systems Best Paper Award, and the Best Paper Award at the 2019 IEEE Artificial Intelligence Circuits and Systems Conference for his work in neuromorphic vision.
\end{IEEEbiography}

\vspace{-2cm}

\begin{IEEEbiography}[{\includegraphics[width=1in,height=1.2in,clip,keepaspectratio]{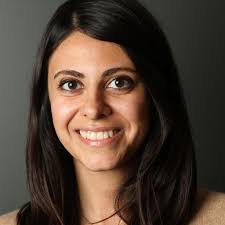}}]{Melika Payvand} 
    (M'19)  is a research scientist at the Institute of Neuroinformatics, University of Zurich and ETH Zurich. She received her M.S. and Ph.D. degree in electrical and computer engineering from the University of California Santa Barbara in 2012 and 2016 respectively. Her research activities and interest is in exploiting the physics of the computational substrate for online learning and sensory processing.  She is part of the scientific committee of the Capocaccia workshop for neuromorphic intelligence, is serving as a technical committee member of Neural Systems, Applications and Technologies in Circuits and System society and as a technical program committee for International Symposium on Circuits and Systems (ISCAS). She is a guest editor of Frontiers in Neuroscience and is the winner of the best neuromorph award of the 2019 Telluride neuromorphic workshop.
\end{IEEEbiography}

\vspace{-2cm}

\begin{IEEEbiography}[{\includegraphics[width=1in,height=1.2in,clip,keepaspectratio]{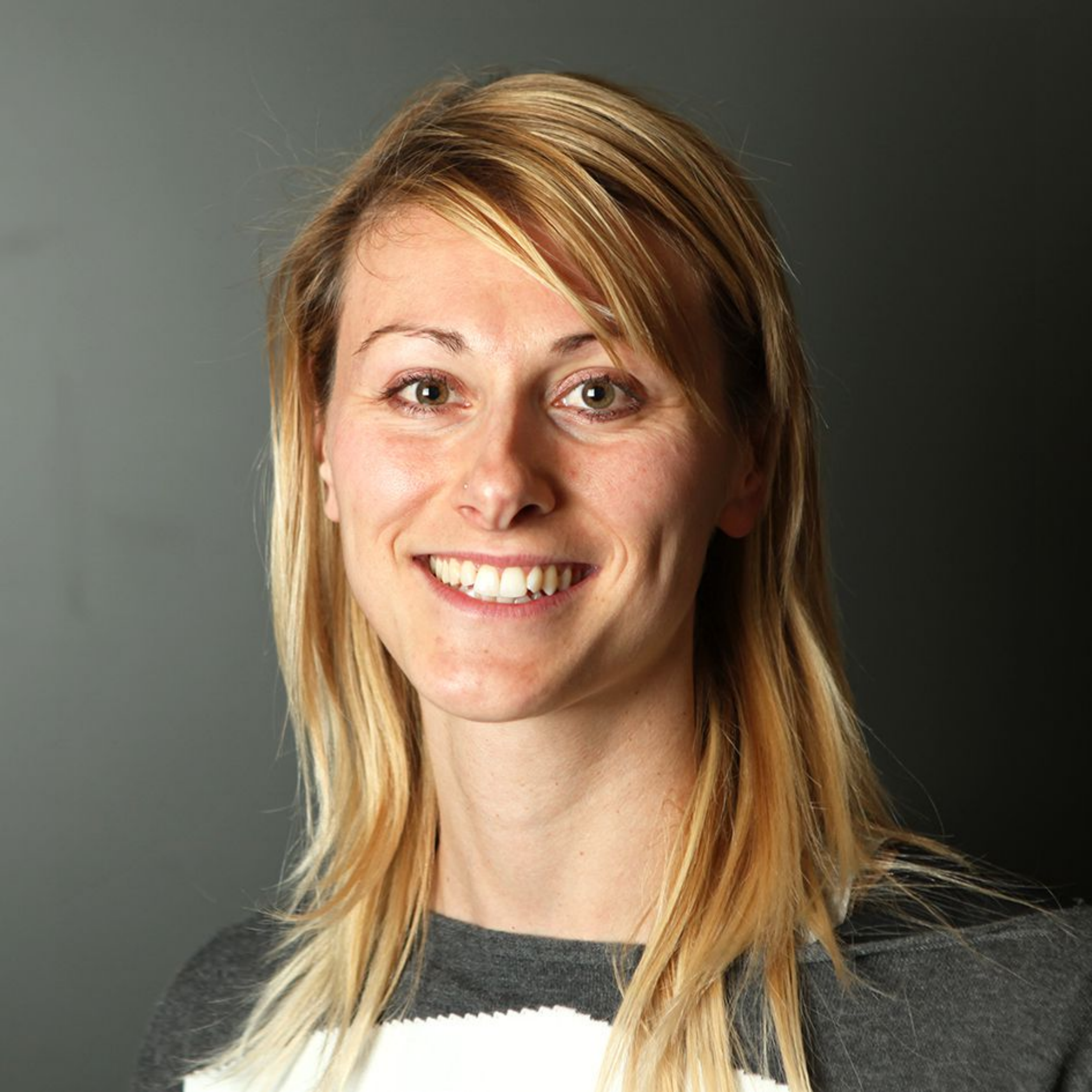}}]{Elisa Donati} 
    (M'14) received the B.Sc, MSc degree in Biomedical Engineering from University of Pisa, Pisa, Italy (cum laude), and the Ph.D. degree in BioRobotics from the Sant'Anna School of Advanced Studies, Pisa, Italy in 2016. Currently, she is a Senior Scientist at the Institute of Neuroinformatics, University of Zurich and ETHZ where she is training as a neuromorphic engineer. Her research interests include how to interface neurorobotics and neuromorphic engineering for building smart and wearable biomedical devices. In particular, she is interested in designing VLSI systems for prosthetic devices, such as adaptive neuromorphic pacemakers. Another recent application includes a neuromorphic processor for controlling upper limb neuroprosthesis. She is investigating how to process EMG data to extract features to produce motor commands by using spiking neural networks. Dr Donati serves as an associate editor of Frontiers in Neuromorphic Engineering and she is a TC member of Neural Systems and Applications of the circuit and system society and of the Biomedical circuit and system society. As member she is part of the commission that is organizing the 2nd IEEE international conference on artificial intelligence circuits and systems.
\end{IEEEbiography}

\newpage
\begin{IEEEbiography}[{\includegraphics[width=1in,height=1.2in,clip,keepaspectratio]{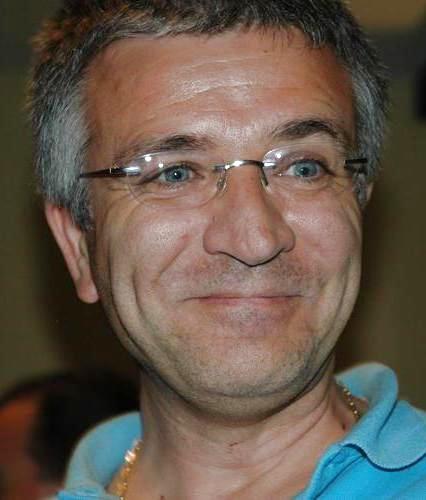}}]{Bernabe Linares-Barranco} 
    (M'90-S'06-F'10) received the B. S. degree in electronic physics in June 1986 and the M. S. degree in microelectronics in September 1987, both from the University of Seville, Sevilla, Spain. From September 1988 until August 1991 he was a Graduate Student at the Dept. of Electrical Engineering of Texas A\&M University. He received a first Ph.D. degree in high-frequency OTA-C oscillator design in June 1990 from the University of Seville, Spain, and a second Ph.D. degree in analog neural network design in December 1991 from Texas A\&M University, College-Station, USA. Since June 1991, he has been a Tenured Scientist at the "Instituto de Microelectrónica deSevilla", (IMSE-CNM-CSIC) Sevilla, Spain, which since 2015 is a Mixed Center between the University of Sevilla and the Spanish Research Council (CSIC). From September 1996 to August 1997, he was on sabbatical stay at the Department of Electrical and Computer Engineering of the Johns Hopkins University. During Spring 2002 he was Visiting Associate Professor at the Electrical Engineering Department of Texas A\&M University, College-Station, USA. In January 2003 he was promoted to Tenured Researcher, and in January 2004 to Full Professor. Since February 2018, he is the Director of the "Insitituto de Microelectrónica de Sevilla". He has been involved with circuit design for telecommunication circuits, VLSI emulators of biological neurons, VLSI neural based pattern recognition systems, hearing aids, precision circuit design for instrumentation equipment, VLSI transistor mismatch parameters characterization, and over the past 25 years has been deeply involved with neuromorphic spiking circuits and systems, with strong emphasis on vision and exploiting nanoscale memristive devices for learning. He is co-founder of two start-ups, Prophesee SA (www.prophesee.ai) and GrAI-Matter-Labs SAS (www.graimatterlabs.ai), both on neuromorphic hardware. He is an IEEE Fellow since January 2010.
\end{IEEEbiography}

\vspace{-8.75cm}

\begin{IEEEbiography}[{\includegraphics[width=1in,height=1.2in,clip,keepaspectratio]{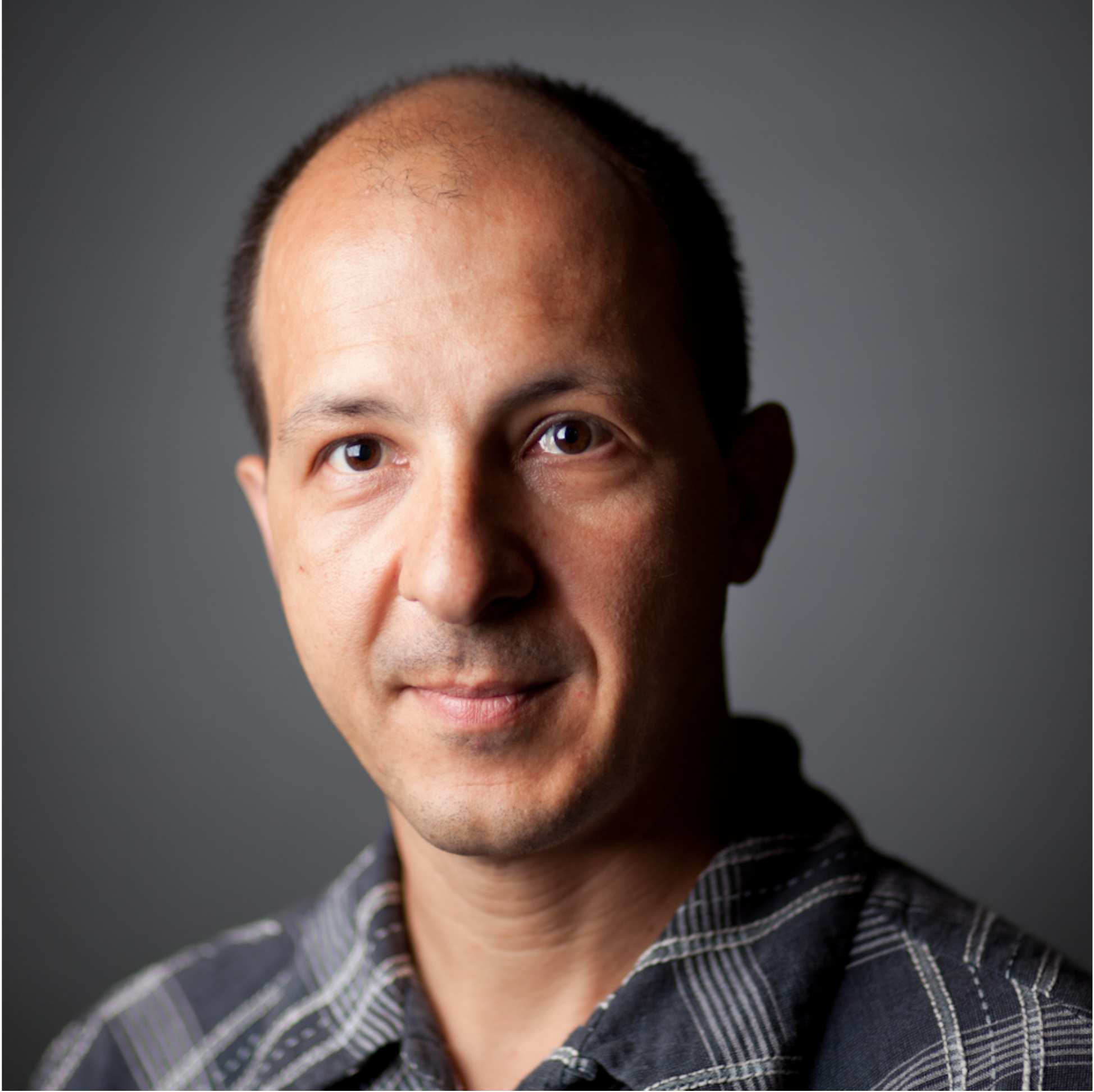}}]{Giacomo Indiveri} 
    (SM'19) is a dual Professor at the Faculty of Science of the University of Zurich and at Department of Information Technology and Electrical Engineering of ETH Zurich, Switzerland. He is the director of the Institute of Neuroinformatics (INI) of the University of Zurich and ETH Zurich. He obtained an M.Sc. degree in electrical engineering and a Ph.D. degree in computer science from the University of Genoa, Italy. He was a post-doctoral research fellow in the Division of Biology at Caltech and at the Institute of Neuroinformatics of the University of Zurich and ETH Zurich. He was awarded an ERC Starting Grant on "Neuromorphic processors" in 2011 and an ERC Consolidator Grant on neuromorphic cognitive agents in 2016. His research interests lie in the study of neural computation, with a particular focus on spike-based learning and selective attention mechanisms. His research and development activities focus on the full custom hardware implementation of real-time sensory-motor systems using analog/digital neuromorphic circuits and emerging memory technologies. 
\end{IEEEbiography}

\end{document}

%% file: acronym.tex
\acrodef{EMG}[EMG]{electromyography}
\acrodef{EEG}[EEG]{electroencephalography}
\acrodef{ECG}[ECG]{electrocardiography}
\acrodef{HFO}[HFO]{High-Frequency Oscillations}
\acrodef{LFP}[LFP]{Local Field Potential}
\acrodef{BMI}[BMI]{Brain-Machine Interface}
\acrodef{SNN}[SNN]{Spiking Neural Network}
\acrodef{RNN}[RNN]{Recurrent Neural Network}
\acrodef{SRNN}[SRNN]{Spiking Recurrent Neural Network}
\acrodef{CNN}[CNN]{Convolutional Neural Network}
\acrodef{AER}[AER]{Address Event Representation}
\acrodef{ANN}[ANN]{Artificial Neural Network}
\acrodef{FPGA}[FPGA]{Field Programmable Gate Array}
\acrodef{PCA}[PCA]{Principle Component Analysis}
\acrodef{SP}[SP]{Separation Property}
\acrodef{ELM}[ELM]{Extreme Learning Machine}
\acrodef{DPI}[DPI]{Differential Pair Integrator}
\acrodef{SVM}[SVM]{Support Vector Machine}
\acrodef{WTA}[WTA]{Winner-take-all}
\acrodef{TPCA}[TPCA]{temporal PCA}
\acrodef{TSNE}[TSNE]{t-distributed stochastic neighbor embedding}
\acrodef{DNN}[DNN]{Deep Neural Network}
\acrodef{VMM}[VMM]{Vector Matrix Multiplication}
\acrodef{MAC}[MAC]{multiply-and-accumulate}
\acrodef{AI}[AI]{Artificial Intelligence}
\acrodef{DL}[DL]{Deep Learning}
\acrodef{GPU}[GPU]{Graphics Processing Unit}
\acrodef{ASIC}[ASIC]{Application Specific Integrated Circuit}
\acrodef{TFLOPS}[TFLOPS]{trillion floating-point operations per second}
\acrodef{MLP}[MLP]{Multi-Layer Perceptron}
\acrodef{LSTM}[LSTM]{Long Short Term Memory}
\acrodef{MDNN}[MDNN]{Memristive DNN}
\acrodef{PCM}[PCM]{Phase Change Memory}
\acrodef{RRAM}[RRAM]{Resistive Random Access Memory}
\acrodef{CBRAM}[CBRAM]{Conductive Bridge RAM}
\acrodef{STT-MRAM}[STT-MRAM]{Spin Transfer Torque Magnetic RAM}
\acrodef{ML}[ML]{Machine Learning}
\acrodef{TPU}[TPU]{Tensor Processing Unit}
\acrodef{IoT}[IoT]{Internet of Things}
\acrodef{PoC}[PoC]{Point of Care}
\acrodef{RNN}[RNN]{Recurent Neural Network}
\acrodef{IoMT}[IoMT]{Internet of Medical Things}
\acrodef{ECog}[ECog]{Echocardiogram}
\acrodef{BCI}[BCI]{Brain Computer Interface}
\acrodef{DVS}[DVS]{Dynamic Vision Sensor}
\acrodef{APS}[APS]{Active Pixel Sensor}
\acrodef{EDP}[EDP]{Energy-Delay Product}
\acrodef{MRAM}[MRAM]{Magnetoresistive Random Access Memory}
\acrodef{PWM}[PWM]{Pulse Width Modulator}
\acrodef{FPAA}[FPAA]{Field Programmable Analog Array}
\acrodef{CAB}[CBA]{Computational Analog Block}
\acrodef{ADC}[ADC]{Analog-to-digital Converter}
\acrodef{DAC}[DAC]{Digital-to-analog Converter}
\acrodef{STDP}[STDP]{Spike-timing-dependent plasticity}
\acrodef{LIF}[LIF]{Leaky Integrate and Fire}
\acrodef{SDSP}[SDSP]{Spike Driven Synaptic Plasticity}
\acrodef{PPG}[PPG]{Photoplethysmography}
\acrodef{BMI}[BMI]{Brain Machine Interface}
\acrodef{PSM}[PSM]{Patient-specific Modeling}
\acrodef{ATC}[ATC]{Asynchronous Temporal-contrast}